\documentclass[aps,pre,superscriptaddress,floatfix]{revtex4-2}
\usepackage{amsmath}
\usepackage{graphicx}
\usepackage{subfigure}
\usepackage{dcolumn}
\usepackage{epstopdf}
\usepackage{natbib}
\usepackage{stfloats}
\usepackage{epsfig,amssymb,subfigure,bm,dsfont}
\usepackage{lipsum} 
\usepackage{appendix}
\usepackage{float}
\usepackage{tikz}
\usetikzlibrary{arrows,shapes,chains}

\usepackage[colorlinks, bookmarks=true,breaklinks=true,linkcolor=red, citecolor=blue, linktocpage=true, urlcolor=blue]{hyperref}
\renewcommand\appendix{\par
	\setcounter{section}{0}
	\setcounter{subsection}{0}
	\gdef\thesubsection{\Roman{subsection}}}
\begin{document}

	\preprint{APS/123-QED}
	\title{Wielding Intermittency with Cycle Expansions}
	\author{Huanyu Cao}
	\affiliation{School of science, Beijing University of Posts and Telecommunications, Beijing 100876, China}
	\author{Ang Gao}
	\affiliation{School of science, Beijing University of Posts and Telecommunications, Beijing 100876, China}
	\author{Haotian Zheng}
	\affiliation{School of science, Beijing University of Posts and Telecommunications, Beijing 100876, China}
	\author{Yueheng Lan}
	\email{lanyh@bupt.edu.cn}
	\affiliation{School of science, Beijing University of Posts and Telecommunications, Beijing 100876, China}
	\affiliation{State Key Lab of Information Photonics and Optical Communications, Beijing University of Posts and Telecommunications, Beijing 100876, China}
	
	\begin{abstract}
		\noindent\textbf{Abstract.}  As periodic orbit theory works badly on computing the observable averages of dynamical systems with intermittency, we propose a scheme to cooperate with cycle expansion and perturbation theory so that we can deal with intermittent systems and compute the averages more precisely. Periodic orbit theory assumes that the shortest unstable periodic orbits build the framework of the system and provides cycles expansion to compute dynamical quantities based on them, while the perturbation theory can locally analyze the structure of dynamical systems. The dynamical averages may be obtained more precisely by combining the two techniques together. Based on the integrability near the marginal orbits and the hyperbolicity in the part away from the singularities in intermittent systems, the chief idea of this paper is to revise intermittent maps and maintain the natural measure produced by the original maps. We get the natural measure near the singularity through the Taylor expansions and periodic orbit theory captures the natural measure in the other parts of the phase space. We try this method on 1-dimensional intermittent maps with single singularity, and more precise results are achieved.
	\end{abstract}
	\maketitle

	\section{\label{sec:1}Introduction}
		From the perspective of statistical physics, in chaotic systems the long-time evolution of typical initial conditions leads to the same asymptotic distribution, the so-called natural measure, which is invariant and effectively used in computing averages of physical observables~\cite{lan2010cycle}. Due to the intrinsic difficulty of depicting the natural measure precisely at different physical states based on system evolution equations~\cite{hao1990chaos}, great challenges are still present in an accurate computation of physical averages in non-equilibrium systems. Fortunately, the strange attractor is densely covered by unstable periodic orbits (UPOs), which could be conveniently used to compute these averages instead of the natural measure. Periodic orbit theory (POT)~\cite{artuso1990recycling1,cvitanovic2005chaos,auerbach1987exploring,cvitanovic1999spectrum,cvitanovic1991periodic} relates global averages to the eigenvalues of appropriate evolution operators~\cite{cvitanovic2005chaos} which is powerful for reliable and accurate analysis in nonlinear chaotic systems. 
		
    	The method of cycle expansions~\cite{artuso1990recycling1,cvitanovic2005chaos,cvitanovic1988invariant,cvitanovic1991periodic} utilizes short periodic orbits (cycles) to compute the spectrum or specifically the leading eigenvalue.
		For uniform hyperbolic systems, cycle expansions converge exponentially or even super-exponentially~\cite{cvitanovic2005chaos}. However, for non-hyperbolic systems, the convergence is considerably slowed down~\cite{artuso1990recycling2}, which casts a shadow over possible application of periodic orbit theory and cycle expansions. Non-hyperbolicity is a very general aspect of dynamics generated near critical points or marginally stable phase space regions in a dynamical system. For example, in fluid flows, one often observes long periods of regular dynamics (laminar phases) interrupted by irregular chaotic bursts, with the distribution of laminar phase intervals well described by a power law~\cite{chate1987transition,chate1994spatiotemporal}. This phenomenon of alternating motion between a chaotic and a regular region in the phase space is called intermittency~\cite{manneville1979intermittency,pomeau1980intermittent,devaney1989introduction}, which invariably leads to drastic dynamical consequences, such as singular natural measures in general, or localisation and quasi-regular eigenfunctions in semi-classical considerations~\cite{tanner1995semiclassical,tanner1996semiclassical,tanner1997chaotic}. 
		
    	When the symbolic dynamics of a particular system is unknown or the hyperbolicity is not so uniform, cycle expansions by stability ordering~\cite{dettmann1997stability} could be a choice. For some stability cutoffs, the cycle expansion converges exponentially with cycle length in some regime and usually faster than that based purely on the topological length of the flow. Nevertheless, some extremely long cycles may need to be involved in the expansion with stability ordering, and an excessively large cutoff may lead to harmful result. Hence the results of stability-ordered expansions should always be tested for robustness by varying the stability cutoff and the size of the cycle set, which will greatly increase the amount of calculation. Besides, several schemes to accelerate the expansion based on analyticity of the spectral functions have been proposed. The tail resummation technique~\cite{artuso1990recycling2} tries to expand the radius of convergence of cycle expansion by removing leading poles of the zeta function. However, in the presence of marginally stable cycles, the tail contribution is led only by a logarithmic convergence thus limiting the accuracy of results. In \cite{artuso2003cycle}, R. Artuso and P. Cvitanovi\'c et al. discuss the properties of the Perron-Fr{\" o}benius operator for intermittent systems, and show that intermittency induces branch cuts in dynamical zeta functions. Marginally stable orbits are incorporated into cycle expansions indirectly through infinite sums over infinitely many symbolic regions. The fundamental term determines the main structure of the zeta function in terms of the leading order branch cut and can be evaluated by constructing an analytic continuation through resummation or by integral transformations. However, how to analytically express curvature contributions to extract the detailed information buried in the intermittent dynamics is faced with difficulty.
		
		In this paper, we discuss an alternative way to deal with intermittency. The marginally stable orbits could be avoided by punching a role around where the non-hyperbolicity is treated with analytic approximation and cycle expansions are implemented outside this region, thus retaining the handsome exponential convergence. The key point is to revise the intermittent map but keep invariant the natural measure as explained in the following. First, the phase space is separated into a hyperbolic and a non-hyperbolic part. Second, a local approximation of the natural measure in the non-hyperbolic region is derived through a Taylor expansion while cycle expansions compute the measure in the hyperbolic part. Finally, a match of the flow out and in could be used to estimate the exchange of measure between the two parts.  The new treatment seems much more accurate than a direct application of the cycle expansion and thus much subdues the trouble brought by intermittency. 
		
		After a brief review of POT in Sect.~\ref{sec:POT}, we describe the difficulty associated with intermittency in cycle expansions in Sect.~\ref{sec:2}. In Sect.~\ref{sec:3}, we introduce a modification of the dynamics, using the idea of induced map \cite{prellberg1992maps} to keep the natural measure while gaining hyperbolicity. Details of our scheme for computing averages of observables are discussed in Sect.~\ref{sec:4}. In Sect.~\ref{sec:5}, several examples are used to demonstrate the validity of the new technique in 1-d intermittent cases. Finally in Sect.~\ref{sec:6}, we summarize the paper and point out possible directions for future investigation. 

	\section{\label{sec:POT}Periodic Orbit Theory}
		With physical intuition and mathematical rigor, the POT puts an alternative way to grasp the main characteristics of nonlinear dynamics and supplies a formalism relating phase space averages to periodic orbits, which carry both topological and dynamics information.
		
		Usually, the average of an observable $a(x)$ is evaluated~\cite{cvitanovic2005chaos} along a trajectory from a typical initial point $x_0$ in phase space $\mathcal{M}$
		\begin{eqnarray}{\label{Formula:timeaverage}}
			\bar{a}_{x_0}=\lim\limits_{n\to\infty}\frac{A^n}{n}=\lim\limits_{n\to\infty}\frac{1}{n}\sum_{i=0}^{n-1}a(f^{i}(x_0))
			\,,
		\end{eqnarray}
		where $x_{i+1}=f(x_i)$ is the given map describing the dynamics and $A^n(x_0)=\sum_{i=0}^{n-1}a(f^{i}(x_0))$ is the integrated observable~\cite{cvitanovic2005chaos}. In this way, a time average may be computed with convenience but the convergence could be very slow sometimes or even not reliable at all occasionally in the presence of non-hyberbolicity. If a measure $\rho(x)$ exists in the phase space another average could be defined 
		\begin{eqnarray}{\label{Formula:phaseaverage}}
			\langle a\rangle_{\rho}=\int_{\mathcal{M}}a(x)\rho(x)dx
			\,,
		\end{eqnarray}
		which is of course dependent on $\rho(x)$. If the dynamics is ergodic, a natural measure $\rho_0(x)$ exists for which the two averages $\langle a\rangle_{\rho_0}=\bar{a}_{x_0}$ for almost all $x_0$. Sometimes, for convenience we write $\langle a\rangle_{\rho_0}=\langle a\rangle$.  Nevertheless, in the phase space of a typical nonlinear system, $\rho_0(x)$ is a singular function defined on a fractal set, which is hard to compute or describe.  

		Interestingly, an alternative approach based on UPOs could be designed to directly compute the averages even without an explicit evaluation of the natural measure. As a strange attractor is densely covered by UPOs, natural measure $\rho_0(x)$ are fully captured by them. Dynamical features can be extracted from the set of UPOs through the evolution operator $\mathcal{L}^n$~\cite{cvitanovic2005chaos}, which evolves the density function $\rho(x)$ according to
		\begin{eqnarray}{\label{Formula:evolution operator}}
			\mathcal{L}^n\circ\rho(x)=\int_{\mathcal{M}}dy\delta(x-f^n(y))e^{\beta A^n(y)}\rho(y)
			\,,
		\end{eqnarray}
		where $n=1,2,\cdots$ for discrete mappings. The kernel function $\mathcal{L}^n(x,y)=\delta(x-f^n(y))e^{\beta A^n}$ depends on the integrated quantity $A^n$ and an auxiliary variable $\beta$. If we set $\beta=0$, $\mathcal{L}$ is the famous Perron-Fr{\" o}benius operator~\cite{cvitanovic2005chaos}. 

		Denoting the spectrum of $\mathcal{L}$ by $\{s_m\}_{m\in\mathbb{N}}$ with $\mathrm{Re}(s_m)>\mathrm{Re}(s_{m+1})$. If the linear operator $\mathcal{L}$ can be thought of as a matrix, high powers of which are dominated by the largest eigenvalue, specifically
		\begin{eqnarray}{\label{Formula:trace approximation}}
			\mathcal{L}^n I(x) =\sum_{m}b_m \phi_m(x) e^{ns_m}\sim b_0 \phi_0(x) e^{ns_0}, n\to\infty
			\,,
		\end{eqnarray} 
		where $I(x)\equiv 1$ is the identity function and expressed as an expansion of the eigenfunctions $\phi_m(x)$ of $\mathcal{L}$, {\em i.e.}, $I(x)=\sum_m b_m \phi_m(x)$. Thus, in terms of the evolution operator, we have 
		\begin{eqnarray}{\label{Formula:e beta n}}
			\langle e^{\beta A^n}\rangle_{I(x)}=\int_{\mathcal{M}}dx \mathcal{L}^n\circ I(x)\sim e^{ns_0}
			\,,
		\end{eqnarray} 
		as $n\to \infty$, where $s_0$ is the leading eigenvalue which is a function of $\beta$. Thus 
		\begin{equation}
 			s_0(\beta)=\lim\limits_{n\to\infty}\frac{1}{n}\ln \langle e^{\beta A^n}\rangle_{I(x)} 
 			\,.\label{eq:s0}
		\end{equation}
		Thus, if the system is ergodic, the average
		\begin{equation}
			\langle a \rangle=\lim\limits_{n\to\infty}\frac{1}{n} \frac{\langle  A^n\rangle_{I(x)}}{\langle  1\rangle_{I(x)}}=\frac{d s_0(\beta)}{d \beta}_{\beta=0}
		\end{equation}
		is directly related to the leading eigenvalue. So, all we need to do is extract the spectrum of $\mathcal{L}$, especially the leading one. Through the identity between the determinant and trace of an arbitrary square matrix $M$ in the matrix algebra: $\mathrm{det}(M)=\mathrm{exp}(\mathrm{tr}\,\mathrm{ln}(M))$, the spectrum of the linear operator $\mathcal{L}$ is determined by solving the resolvent equation $\mathrm{det}(\mathbf{1}-z\mathcal{L})=0$ from the set of UPOs through the trace \cite{cvitanovic2005chaos,artuso1990recycling2}
		\begin{align}{\label{Formula:trace formula}}
			\mathrm{tr}(\mathcal{L}^n)=&\int_{\mathcal{M}}dx\mathcal{L}^n(x,x)=\int_{\mathcal{M}}dx\delta(x-f^n(x))e^{\beta A^n}\nonumber\\
			=&\sum_{f^n(x_i)=x_i}\frac{e^{\beta A^n(x_i)}}{|\mathrm{det}(\mathbf{1}-M_n(x_i))|},\forall n\in\mathbb{Z}^+
			\,,
		\end{align}
		where $x_i$ is a periodic point of period $n$ and $M_n(x_i)$ is the Jacobian matrix of $f^n(x)$ evaluated at $x_i$.  Thus the spectral determinant~\cite{cvitanovic2005chaos,artuso1990recycling1,artuso1990recycling2} is 
		\begin{align}{\label{Formula:spectral determinant}}
			\mathrm{det}(\mathbf{1}-z\mathcal{L})&=\mathrm{exp}(\mathrm{tr}\,\mathrm{ln}(\mathbf{1}-z\mathcal{L}))=\mathrm{exp}\left(-\sum_{n=1}^{\infty}\frac{z^n}{n}\mathrm{tr}(\mathcal{L}^n)\right)\nonumber\\
			&=\mathrm{exp}(-\sum_{p}\sum_{r=1}^{\infty}\frac{1}{r}\frac{z^{n_pr}e^{r\beta A_p}}{|\mathrm{det}(\mathbf{1}-\mathit{M}_p^r)|})
			\,,
		\end{align}
		where $p$ denotes prime cycles which are not repeats of shorter ones and $n_p$ is the length of the cycle $p$. $A_p$ and $M_p$ are the integrated physical quantity and the Jacobian along the prime cycle $p$. If the system is hyperbolic, we can make the approximation
		\begin{eqnarray}{\label{Formula:stability approximation}}
			\frac{1}{|\mathrm{det}(\mathbf{1}-M_p^r)|}\approx\frac{1}{|\Lambda_p|^r},
		\end{eqnarray}
		where $\Lambda_p=\prod_{e}\Lambda_{p,e}$ is the product of expanding eigenvalues of the matrix $M_p$. With $r\to\infty$, the spectral determinant Eq.~(\ref{Formula:spectral determinant}) becomes the dynamical zeta function~\cite{cvitanovic2005chaos,cvitanovic1999spectrum}
		\begin{eqnarray}{\label{Formula:dynamical zeta function}}
			\frac{1}{\zeta}=\prod_{p}(1-t_p),
		\end{eqnarray}
		where $t_p=\frac{z^{n_p}e^{\beta A_p}}{|\Lambda_p|}$. The dynamical zeta function is the $0$th-order approximation of the spectral determinant and they have identical leading eigenvalue but different analytic properties~\cite{cvitanovic2005chaos}. 
		
\begin{figure}[H]	
			\centering
			\subfigure[]{\includegraphics[scale=0.55]{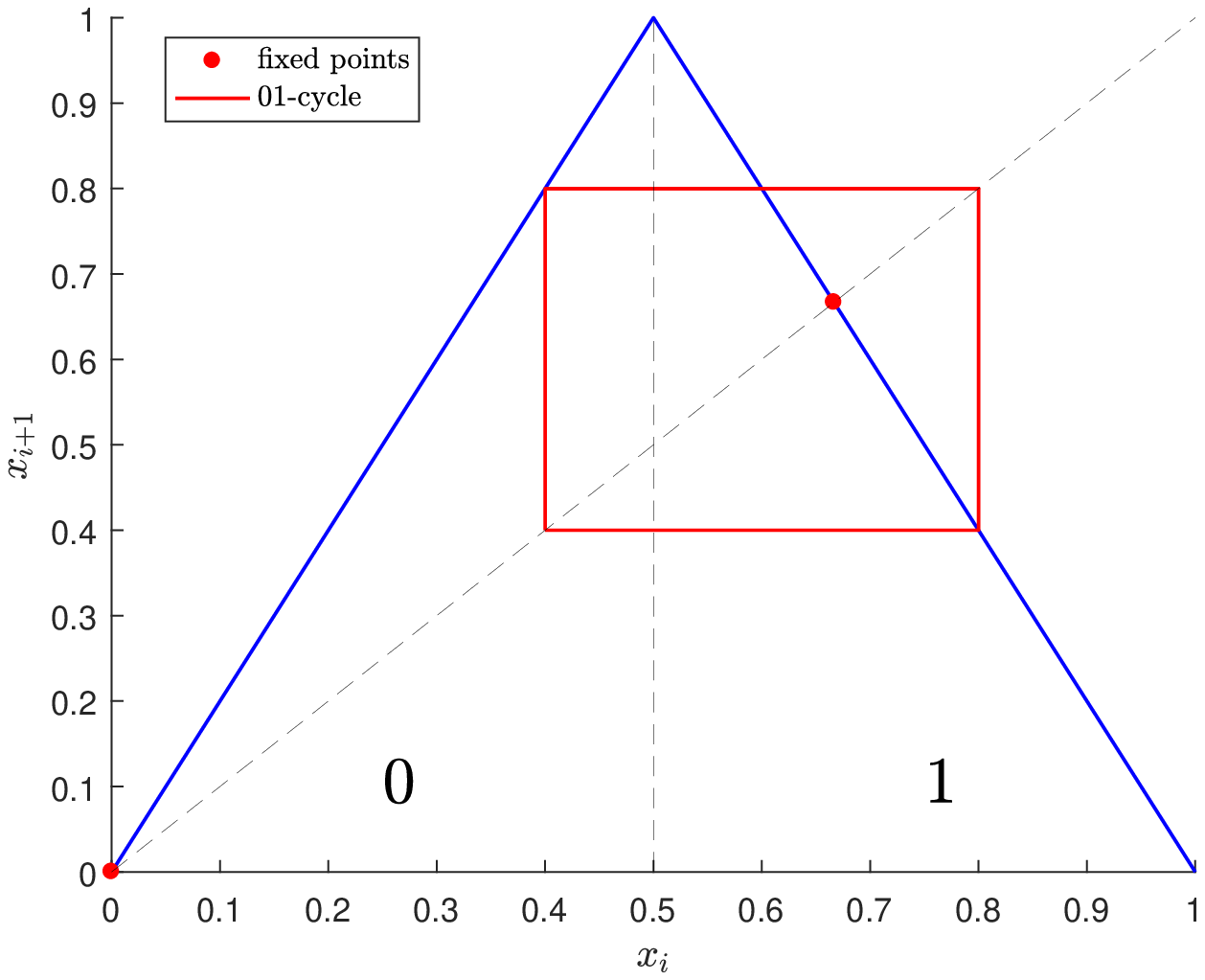}}
			\subfigure[]{\includegraphics[scale=0.55]{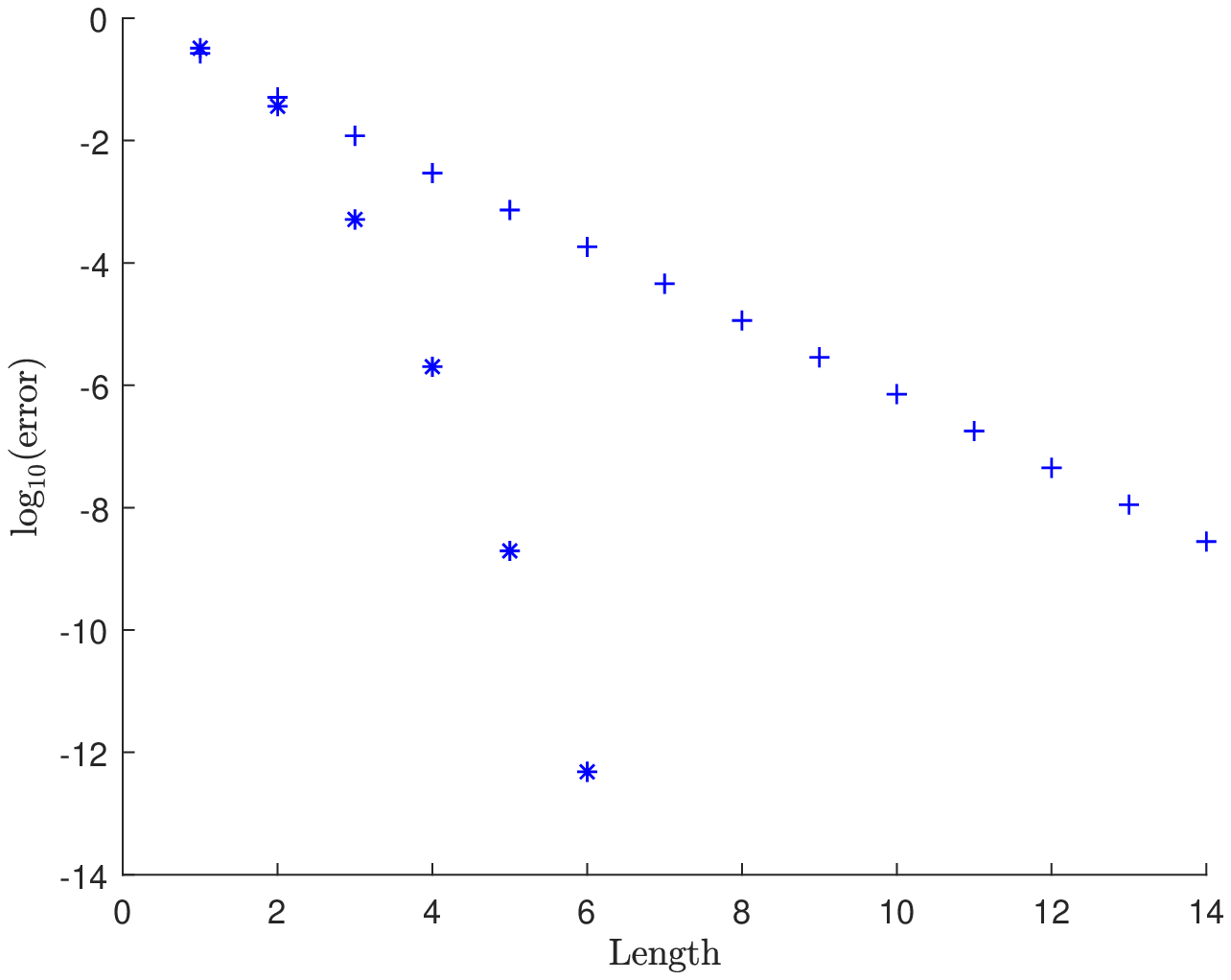}}
			\caption{(a) The tent map $f$ on the unit interval $[0,1]$ with its symbolic regions denoted by $0$ and $1$. The fixed points $x=0$ and $x=2/3$ are actually period one orbits and represented by $\bar{0}$ and $\bar{1}$ respectively. The period two orbit has the sequence $0101...$ and is denoted by $\bar{01}$. (b) The convergence rates of cycle expansions are marked with the decay of the computation errors with the truncation length, {\em e.g.},  in the estimation of the average $\langle x\rangle$ ($+$, exponential decay) or the escape rate ($*$, super-exponential decay). The values obtained with the largest truncation length are considered as the “exact” values in the estimation of errors.}
			\label{graphic:tent map}
		\end{figure}

		In order to find all the periodic orbits, we have to resort to symbolic dynamics which encodes all the possible orbits~\cite{lan2010cycle}. As an example, let's consider the tent map
		\begin{eqnarray}{\label{Formula:tent map}}
			x \mapsto f(x)=\begin{cases}
			f_0(x)=2x&x\in[0,\frac{1}{2})\\
			f_1(x)=2-2x&x\in[\frac{1}{2},1]
			\end{cases},
		\end{eqnarray}
		which is uniformly hyperbolic and has a critical point at $x=1/2$. We label the two non-overlapping intervals $[0,1/2)$ and $[1/2,1]$ with ``$0$'' and ``$1$'' respectively so that a trajectory is uniquely associated with a binary symbol sequence $s_0s_1s_2s_3..., s_i\in\{0,1\}$ called itinerary according to the intervals which the orbit consecutively visits. A family of orbits are denoted as $s_0s_1s_2...s_k$, which start and visit the same intervals within $k$ iterations. For example, the period two orbit in Fig.~\ref{graphic:tent map}(a) is described by the infinite sequence $010101...$, which may be denoted as $\bar{01}$ and has a topological length of $2$. 
		Utilizing the property of orbit shadowing in nonlinear dynamics~\cite{cvitanovic2005chaos}, cycle expansion is often an efficient method for computing the spectrum, with short periodic orbits capture the major part of the natural measure and longer cycles deliver systematic curvature corrections. For maps with binary symbolic dynamics, Eq.~(\ref{Formula:dynamical zeta function}) can be expanded as
		\begin{eqnarray}{\label{Formula:cycle expansion}}
			\begin{split}
				\frac{1}{\zeta}=&1-\sum_{f}t_f-\sum_{p}c_p=1-t_0-t_1-[(t_{01}-t_0t_1)]\\
				-&[(t_{001}-t_{01}t_0)+(t_{011}-t_{01}t_1)]-...,
			\end{split}
		\end{eqnarray}
		where the fundamental terms $t_f$ include all unbalanced, not shadowed prime cycles and the rest terms $c_p$, called curvature corrections, consist of longer prime cycles and pseudo-cycles that shadow them. Cycle expansions are dominated by fundamental terms, with long orbits shadowed by short orbits, so that curvature corrections decay exponentially or even super-exponentially (shown in Fig.~\ref{graphic:tent map}(b)) if uniform hyperbolicity is assumed. The cancellation between prime cycles and pseudo-cycles reflects the smoothness of the underlying dynamics. 
		
		As there often exist infinitely many unstable cycles in a chaotic system, truncation to the spectral functions is also necessary in practical computation. The usually adopted truncation with cycle length corresponds to a geometric envelope approximation of the original map~\cite{cvitanovic1991periodic2} and higher order truncations lead to a more accurate evaluation as shown in Fig.~\ref{graphic:tent map}(b). 
		
	\section{\label{sec:2} Difficulty Associated with Intermittency}
		When a dynamical system is uniformly hyperbolic and its symbolic dynamics is a subshift of finite type, it can be shown that the evolution operator~$\mathcal{L}$ has discrete spectra $\{s_m\}_{m\in\mathbb{N}}$ and the associated spectral determinants are entire functions~\cite{ruelle2004thermodynamic,rugh1992correlation} whose zeros $\{z_m\}_{m\in\mathbb{N}}$ are related to the eigenvalues via $z_{\alpha}=e^{-s_{\alpha}}$. The average $\langle a\rangle$ or other dynamical properties can be conveniently computed based on the leading eigenvalue $s_0$ of $\mathcal{L}$ through spectral determinants or dynamical zeta functions. With intermittency, the spectra of evolution operators, however, become continuous and dynamical zeta functions exhibit branch cuts~\cite{cvitanovic2005chaos,artuso2003cycle}. Cycle stabilities grow no longer exponentially with length, but as power laws which leads to poor cancellations, sending cycle expansions awry. 
		
		The difficulty of intermittency comes from the marginally unstable periodic orbit embedded in the phase space. Once the state of the system comes close enough to the marginal orbit, it will stick around for a long time. So not surprisingly, there will be singularities in natural measure, being located near the marginal orbit. The stability eigenvalues of the UPOs near the marginal region are very close or equal to $1$, breaking the exponential instability with respect to the cycle length and rendering the spectral determinant singular. Here is an example of intermittent map 
		\begin{eqnarray}{\label{Formula:2.1}}
			x \mapsto f(x)=\begin{cases}
			f_0(x)=x/(1-\sqrt{x/2})&x\in\mathcal{M}_0=[0,\frac{1}{2})\\
			f_1(x)=2-2x&x\in\mathcal{M}_1=[\frac{1}{2},1]
			\end{cases},
		\end{eqnarray}
		which has a marginally unstable fixed point $x=0$, leading to a natural measure singularity at the origin (Fig.~\ref{graphic:examplemap}(b)). $f_0$ and $f_1$ correspond to the two branches of the map. In this paper, we shall restrict our consideration to $1$-dimensional maps which are similar to $f(x)$, with the marginally unstable fixed point $x=0$ and the approximation
		\begin{eqnarray}{\label{Formula:2.2}}
			x\longmapsto f(x)\sim x+cx^{1+s}
		\end{eqnarray}
		in its neighborhood, which are characterized by the intermittency exponent $s$. In this kind of maps, \(x=0\) is a singularity of the natural measure which takes the form \(\rho\sim\frac{1}{x^s}\) near $x = 0$, to be shown later. If we further assume that this kind of maps have complete binary symbolic dynamics, the eigenvalue of the orbit \(10^{m-1}\) grows algebraically with its length: \(\Lambda_{10^{m-1}}\sim m^{1+1/s}\), when $m$ is large enough~\cite{artuso2003cycle}. This quite makes cycle expansions converge algebraically. 
		
		Nevertheless, the integrability near the marginal points of intermittent maps gives us inspiration and becomes the foundation and starting point of our work. We can easily take advantage of this property in marginal regions, where the natural measure is computed with perturbation theory in combination with cycle expansions in hyperbolic regions.

	\section{\label{sec:3}Separation and Reconstruction Scheme}
		After fully appreciating the difficulty associated with intermittency, it may seem natural to separate the phase space into two parts. One is the marginal part which contains the marginal orbit and the other is a totally hyperbolic part. Hopefully, the cycle expansion on the hyperbolic part will be accelerated after the marginal orbit is removed. The natural measure on the hyperbolic part should not be changed. Therefore, a subtle manipulation of the dynamics defined on this part is needed. We focus on the map $f$ mentioned above and show that for this purpose a comb structure could be introduced with inspiration from the branch structure in \cite{artuso2003cycle}.

		\subsection{\label{subsec:3.1}A comb structure}
\begin{figure}[H]	
				\centering
				\subfigure[]{\includegraphics[scale=0.55]{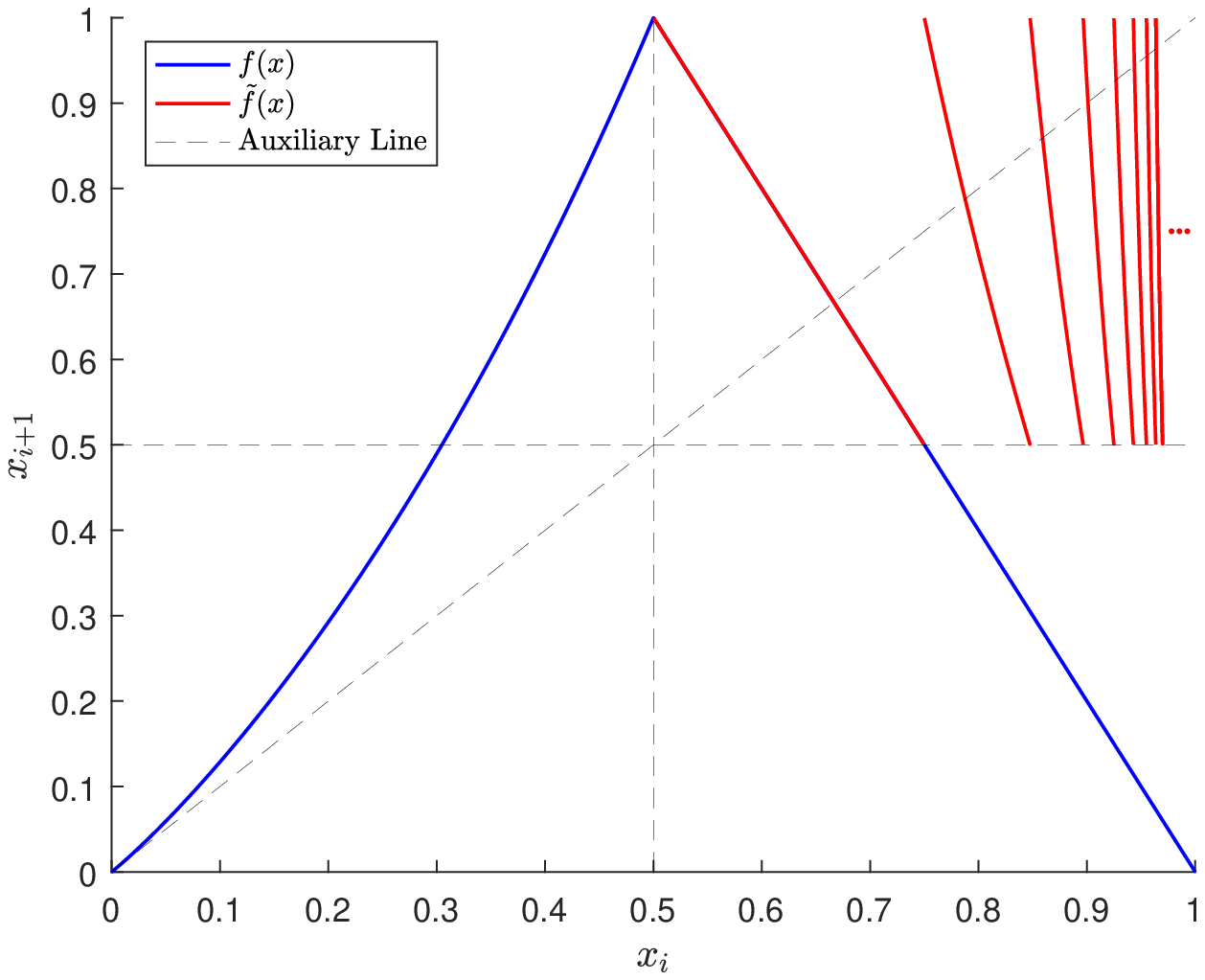}}
				\subfigure[]{\includegraphics[scale=0.55]{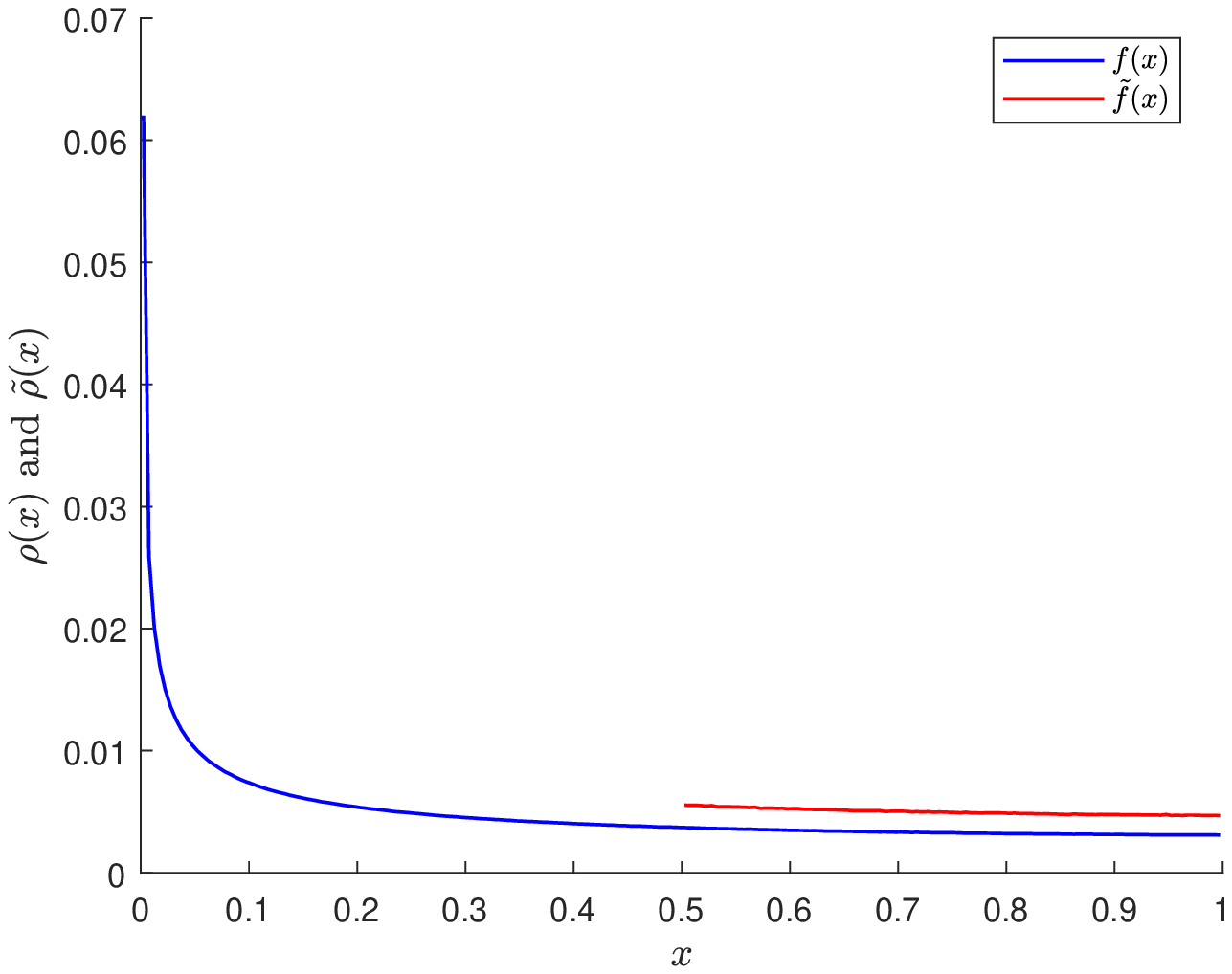}}
				\caption{The intermittency map and the natural measure: (a) The map $f(x)$ (Eq.~(\ref{Formula:2.1}), blue line) and the revised map $\tilde{f}(x)$ (Eq.~(\ref{formula:3.1.1}), red line). The two images coincide partly on the interval $[1/2,1]$ and infinite branches of $\tilde{f}(x)$ on the right are supplied to keep the measure invariant. (b) The natural measure of $f(x)$ (blue line) and $\tilde{f}(x)$ (red line). Obviously, there is a singularity at $x=0$ in the natural measure of $f(x)$.}
				\label{graphic:examplemap}
			\end{figure}

			Our scheme is illustrated for the map $f(x)$ in Eq.~(\ref{Formula:2.1})and certainly applicable to other cases. A convenient and natural way is to just split the two branches of the map, teating $[0,1/2)$ as the marginal part and $[1/2,1]$ as the hyperbolic part. For the hyperbolic part, we construct a comb structure (Fig.~\ref{graphic:examplemap}(a)) to replace the original map but keep the natural measure unchanged. The revised map is defined on $[1/2,1]$ and denoted as $\tilde{f}(x)$ given by
			\begin{small}\begin{eqnarray}{\label{formula:3.1.1}}
				\tilde{f}(x)=\begin{cases}
				f_1(x) & x\in \mathcal{M}_1=[q_1,q_2] \\
				f_2(x)=f_0(f_1(x)) & x\in \mathcal{M}_2=(q_2,q_3] \\
				\vdots & \vdots \\
				f_m(x)=f_0^{m-1}(f_1(x)) & x\in \mathcal{M}_m=(q_m,q_{m+1}] \\
				\vdots & \vdots 
				\end{cases},
			\end{eqnarray}\end{small}
			where $q_1=1/2$ and \(q_m=f_1^{-1}(f_0^{2-m}(q_1)),m\geq2\). The revised map $\tilde{f}(x)$ has a nice property that \(\tilde{f}_m(\mathcal{M}_m)\subset[1/2,1]\), which indicates that $\tilde{f}(x)$ is self-contained and enjoys a complete symbolic dynamics. In addition, a trajectory of $f(x)$ is always mapped to that of $\tilde{f}(x)$, just by lumping the points in $[0,1/2)$ to the first point out of this interval along the trajectory. Furthermore, the natural measure produced by $\tilde{f}(x)$ is proportional to that of $f(x)$ on $[1/2,1]$ 
\begin{eqnarray}{\label{formula:3.1.2}}
				\rho(x)=\lambda\,\tilde{\rho}(x),\,x\in [1/2,1],\,0<\lambda<1
				\,,
			\end{eqnarray}	
where the constant $\lambda$ may be computed through cycle expansions. We put the justification and detailed derivation in App.~\ref{appendix:7.1}. As the maps discussed in this paper are all ergodic, the rough picture of the natural measure may be constructed from the histogram of map iterations as shown in Fig.~\ref{graphic:examplemap}(b), where a singularity near the origin is clearly seen. 
		\subsection{\label{subsec:3.2}Redefinition of the observable}
			After the construction of the revised map \(\tilde{f} : [1/2,1] \mapsto [1/2,1]\), which successfully maintains the natural measure, the dynamical averages of $f(x)$ can be easily obtained. We could calculate averages in $[0,1/2)$ and $[1/2,1]$ by analytical approximation and cycle expansion separately. 
			
			Nevertheless, we are able to take one step further. In fact, the analytical approximation for the natural measure on $[0,1/2)$ may be unnecessary, since we may obtain the averages with a redefinition of the observables and restrict the computation to $[1/2,1]$. Denote an observable by $a(x)$ and its average is $\langle a\rangle =\int_\mathcal{M}a(x)\rho(x)dx$, with $\rho(x)$ being the natural measure. We could redefine the observable on $[1/2,1]$ as
			\begin{small}\begin{eqnarray}{\label{formula:3.2.1}}
				\tilde{a}(x)=\begin{cases}
				a(x) & x\in \mathcal{M}_1=[q_1,q_2] \\
				a(x)+a(f_1(x)) & x\in \mathcal{M}_2=(q_2,q_3] \\
				\vdots & \vdots \\
				a(x)+\sum_{n=2}^{m}a((f_0^{n-2}f_1)(x)) & x\in \mathcal{M}_m=(q_m,q_{m+1}] \\
				\vdots & \vdots 
				\end{cases}\,,
			\end{eqnarray}\end{small}
			which implies a beautiful equation for $\tilde{a}(x)$
			\begin{eqnarray}{\label{formula:3.2.2}}
				\langle\tilde{a}\rangle=\int_{1/2}^{1}\tilde{a}(x)\tilde{\rho}(x)dx=\int_{0}^{1}a(x)\frac{\rho(x)}{\lambda}dx=\frac{\langle a\rangle}{\lambda}
				\,,
			\end{eqnarray}
indicating that all the averages are obtainable with just the revised map $\tilde{f}(x)$ (in theory at least). We relegate the justification to App.~\ref{appendix:7.2}.
		\subsection{\label{subsec:3.3}Dynamical zeta function}
			For the trajectories of $f(x)$ and $\tilde{f}(x)$ have one-to-one correspondence, it is convenient to map the symbol sequences 
			\begin{eqnarray}{\label{formula:3.3.1}}
			\begin{split}
			10^{l-1}\to l;\,10^{l-1}10^{m-1}\to lm;\\10^{l-1}10^{m-1}10^{n-1}\to lmn;\,\dots,
			\end{split}
			\end{eqnarray}
			where $m$ stands for the interval $\mathcal{M}_m$, which has been defined along with $\tilde{f}(x)$. The corresponding cycle has the same stability and integrated observable as the original one but the cycle length is reduced in the revised map. For example, if an itinerary of $f(x)$ has a length of $m + n$, where $n$ stands for the number of points in $[0,1/2)$, the reduced itinerary then only has a length of $m$. These relations make the cycle search quite easy since the new map $\tilde{f}(x)$ does not need to be used. All its periodic orbits can be adapted from those of $f(x)$. The dynamical zeta functions for $f(x)$ and $\tilde{f}(x)$ compare as follows	
			\begin{eqnarray}{\label{formula:3.3.2}}
			\begin{split}
			\frac{1}{\zeta_{f}}&=\prod_p(1-t_p),t_p=\frac{e^{\beta A_p}}{|\Lambda_p|}z^{m+n}\\
			\frac{1}{\zeta_{\tilde{f}}}&=\prod_p(1-t_p'),t_p'=\frac{e^{\beta A_p}}{|\Lambda_p|}z^m\\
			\end{split},
			\end{eqnarray}
			from which we can clearly see that we just need a small change to get the dynamical zeta function of $\tilde{f}(x)$. However, the infinite branches of $\tilde{f}(x)$ still have to be treated as explained below. 
		\subsection{\label{sec:4}One convenient approximation}
		Difficulties still exist in a direct application of Eq.~(\ref{formula:3.3.2}) in practice. Firstly, according to Eq.~(\ref{formula:3.3.2}), infinite number of long orbits should be computed which are reducible to short ones in the new map and thus contribute to the fundamental part in the cycle expansion. This manifests as the infinitely many branches of the comb structure and is impossible to achieve numerically. For example, orbit $0^{m-1}1$ with a large enough $m$ will be reduced into a fixed point located on the $m$th branch of the revised map. Secondly, the redefined observable $\tilde{a}(x)$ might be singular at $x=1$, which certainly slows down the convergence of the cycle expansion. As a result, approximations are still needed to further improve the operation and effectiveness of our scheme.
			
		First, the infinite comb structure is replaced with finitely many branches. For the revised map $\tilde{f}(x)$, we replace all the branches $f_m(x)$ ($m\geq n$) with just one linear branch $\hat{f}_n(x)=\frac{q_1-1}{1-q_n}(x-q_n)+1,x\in \mathcal{M}_n=(q_n,1]$. Obviously, $\hat{f}_n(\mathcal{M}_n)\subset[1/2,1]$. Here, $n$ describes the level of the approximation, being more accurate for larger $n$. Nevertheless, under this approximation, the cycles that pass the interval $(q_n,1]$ cannot be obtained directly from the original map $f(x)$, but have to be found individually. Second, we save the averages in the neighborhood of $x=0$ for different treatment which effectively removes the above mentioned singularity. To achieve this, we just need to analytically estimate the natural measure in the interval $[0,f_1(q_n)]$. As a result, in the interval $\mathcal{M}_n$, the observable $\tilde{a}(x)$ could be redefined as $a(x)+\sum_{k}^{}a((f_0^kf_1)(x))$ where $k$ keeps $(f^k_0f_1)(x)\in (f_1(q_n),q_1)$ and thus finite. The detailed discussion is seen in App.~\ref{appendix:lambda}. 
			
			The computation of the average $\langle a\rangle_0$ in the interval $[0,f_1(q_n)]$ calls for an analytic derivation of the natural measure $\rho(x)$. Expectedly, the redefinition of the observable greatly reduces the approximation interval, which enables a local expansion based on the equation
			\begin{small}\begin{eqnarray}{\label{formula:4.2}}
				\rho(x)=\frac{\rho(f_0^{-1}(x))}{|f_0'(f_0^{-1}(x))|}+\frac{\rho(f_1^{-1}(x))}{|f_1'(f_1^{-1}(x))|},x\in[0,1/2)
				\,,
			\end{eqnarray}\end{small}
since the dynamics near $x=0$ is integrable. The detailed derivation of the analytic approximation is given in the examples below and could also be found in the appendix. Finally, the dynamical average we need is obtained from $\langle a\rangle_0$ in the neighborhood of $x=0$ and $\langle\tilde{a}\rangle$ on the hyperbolic part as
			\begin{eqnarray}{\label{formula:final average}}
				\begin{split}
					\langle a\rangle=&\int_{0}^{f_1(q_n)}\rho(x)a(x)dx+\lambda\int_{q_1}^{1}\tilde{\rho}(x)\tilde{a}(x)dx\\
					=&\langle a\rangle_0+\lambda\langle\tilde{a}\rangle.
				\end{split}
			\end{eqnarray}
	\section{\label{sec:5}Several examples}
		We show that dynamical averages for an intermittent map can be effectively computed through a series of operations including separation, reconstruction, redefinition and analytic computation. The revised map behaves much better than the original one in that the marginal orbit is eliminated and cycle expansion is accelerated. We apply what is described above to compute the averages of the following intermittent maps to verify its effectiveness. Before doing that, some details on numerical computation need to be noted. 
		\subsection{\label{subsec:5.0}Details on numerical computation}
		For all degree $n$ of the approximation, we set a truncation length \(L_{max}=10\) for cycle expansions and found out all the prime cycles no longer than $L_{max}$. Based on the state space Markov partition \cite{sinai1968construction}, symbolic dynamics \cite{kitchens2012symbolic,collet2009iterated,metropolis1973finite} and multiple shooting method \cite{auerbach1987exploring} are very effective in cycle searching. Averages obtained with the degree $n$ of the approximation are used as the ``exact'' values when estimating the errors of those with lower degrees. Errors {\em vs} degree $n$ are plotted to show the convergence in the logarithmic scale.
			
		On the other hand, dynamical averages are estimated with efforts by time averaging (Eq.~(\ref{Formula:timeaverage})) and also with a direct application of the dynamical zeta function, to compare with the new treatment. For each example, we start from $10^2$ different random initial points and iterate $10^8$ times each to get $10^2$ dynamical averages along these trajectories, which are then averaged again, with a standard deviation computed to indicate the accuracy.
		\subsection{\label{subsec:5.1} The example model}
\begin{figure}[H]	
				\centering
				\subfigure[]{\includegraphics[scale=0.55]{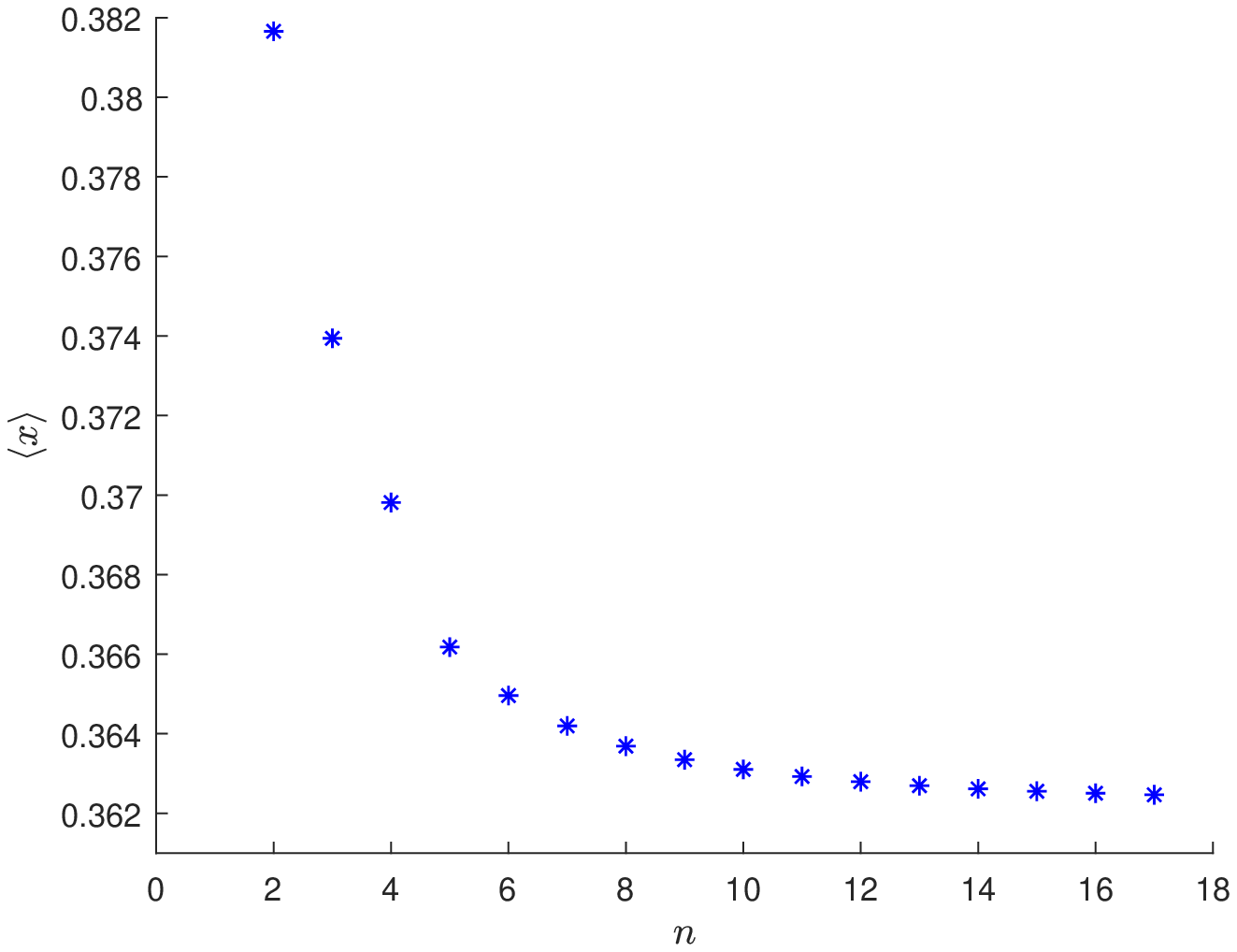}}
				\subfigure[]{\includegraphics[scale=0.55]{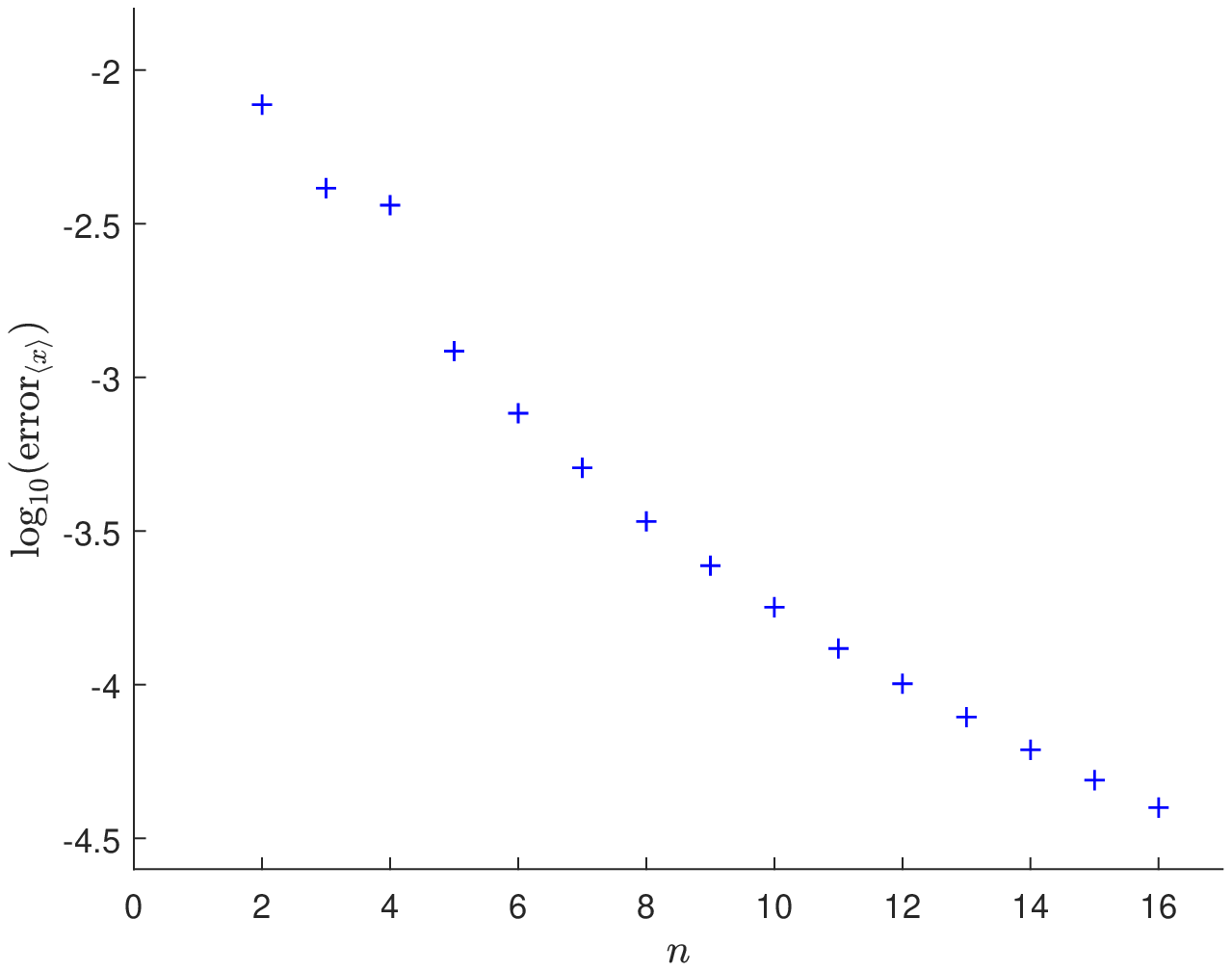}}
				\caption{The change of the average and the error with the approximation level for map Eq.~(\ref{Formula:2.1}) in the new scheme. (a) the average $\langle x\rangle$ becomes more precise as $n$ increases and converges to 0.3624 at $n=17$. (b) the logarithmic error with respect to $n$.}
				\label{graphic:1.1}
			\end{figure}

			We continue with the previous example map $f(x)$ as the first demonstration. Near the marginally unstable fixed point $x = 0$, the map takes the form
			\begin{small}\begin{eqnarray}{\label{formula:5.1.2}}
			\begin{split}
			f(x)=\sum_{i=0}^{\infty}x(\frac{x}{2})^{i/2}=x+\frac{x^{3/2}}{\sqrt{2}}+\mathcal{O}(x^2)
			\end{split}
			\end{eqnarray}\end{small}	 
			where $\mathcal{O}(x^2)$ denotes higher order terms. Obviously, we have the intermittency exponent $s=\frac{1}{2}$ in this map. To simplify the calculation, we repartition the interval $[0,1]$ into the marginal part $[0,f^{-2}_0(q_1))$ and the hyperbolic part $[f^{-2}_0(q_1),1]$. The revised map $\tilde{f}(x)$ is shown in Fig.~\ref{graphic:examplemap}(a) and the redefined observable $\tilde{a}(x)$ are defined as in Eq.~(\ref{formula:3.2.1}), but with the revision described in Sect.~\ref{sec:4}. Based on the new partition, the symbolic dynamics is easily established and the lengths of many prime cycles are much reduced. As mentioned in Sect.~\ref{sec:4}, we merge all the later branches $f_m(x)$, $m\ge n$ into a linear one $\hat{f}_n(x)$. 
			\begin{figure}[H]	
				\centering
					\begin{tikzpicture}[->,>=stealth',shorten >=1pt,auto,node distance=4cm,
					thick,base node/.style={circle,draw,minimum size=16pt}, real node/.style={double,circle,draw,minimum size=35pt}]
					\node[shape=circle,draw=black](-1){-1};
					\node[shape=circle,draw=black](0)[right of=-1]{0};
					\node[shape=circle,draw=black](1)[below of=-1]{1};
					\node[shape=circle,draw=black](m)[right of=1]{m};
					\path[]
					(1) edge [loop below]node {} (1)
					(-1) edge node {} (0)
					(0) edge node {} (1)
					(0) edge node {} (m)
					(m) edge node {} (-1)
					(1) edge node {} (0)
					(1) edge node {} (m)
					(1) edge node {} (-1);
					\end{tikzpicture}
				\caption{The Markov chain of the revised map $\tilde{f}(x)$. It marks the density transportation direction and supports the establishment of symbolic dynamics. The phase space for the hyperbolic dynamics is divided into non-overlapping regions: ``$-1$'', the interval $[f^{-2}_0(q_1),f^{-1}_0(q_1)]$, ``$0$'', $(f^{-1}_0(q_1),q_1]$, ``$1$'', $(q_1,q_2]$ and ``$m$'', the intervals $\mathcal{M}_m$, $m\ge2$.}
				\label{graphic:Markovchain}
			\end{figure}
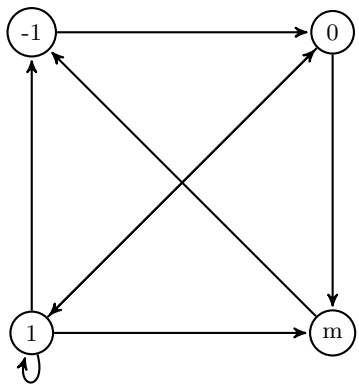
			We try to evaluate the average $\langle x\rangle$ via Eq.~(\ref{formula:final average}). As mentioned above, we need to calculate $\langle \tilde{x}\rangle$ and $\lambda$ through the dynamical zeta function for $\tilde{f}(x)$ in Eq.~(\ref{formula:3.3.2}) and cycle expansion (Eq.~(\ref{Formula:cycle expansion})). The natural measure of $f(x)$ near $x=0$ is analytically approximated as $\rho(x)=\frac{\omega_0}{\sqrt{2}}x^{-1/2}+\frac{\omega_0}{8}+\frac{\omega_0}{32\sqrt{2}}x^{1/2}+\mathcal{O}(x)$ for the computation of $\langle x\rangle_0$ in the interval $[0,f_1(q_n)]$, where $\omega_0$ is the natural measure supported on the interval $(q_n,1]$ (see App.~\ref{appendix:7.3}). The results are shown in Fig.~\ref{graphic:1.1}. Obviously, the average becomes more precise when $n$ is large. $\langle x\rangle$ converges to $0.3624$ at $n=17$ and is closed to $0.3622...$, the time averaging result with the standard deviation equal to $2.457...\times 10^{-4}$. A direct application of the dynamical zeta function gets $0.22$, which is totally off due to the slow convergence incurred by the intermittency. 
		\subsection{\label{subsec:5.2}Bernoulli shift map}
\begin{figure}[H]	
				\centering
				\subfigure[]{\includegraphics[scale=0.55]{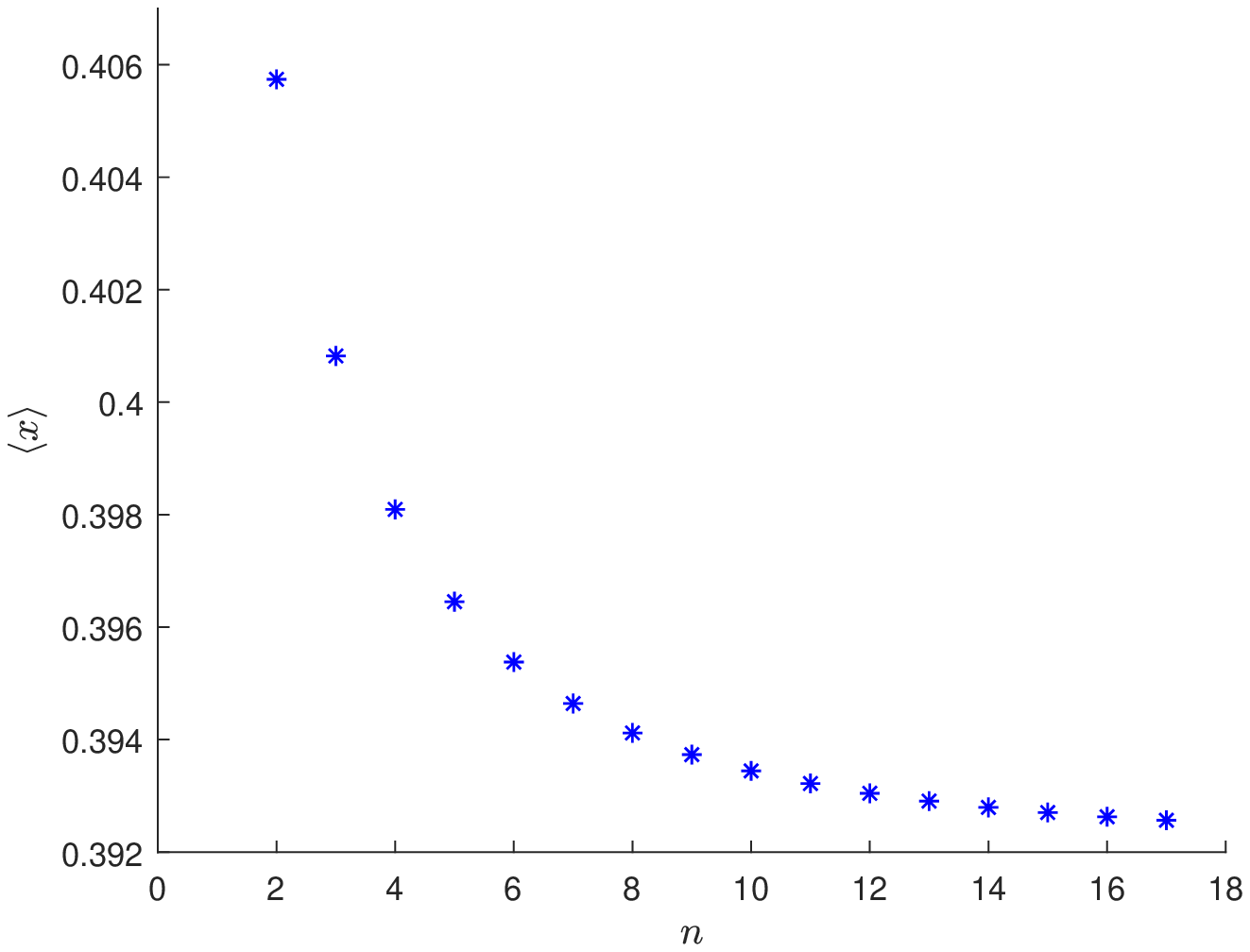}}
				\subfigure[]{\includegraphics[scale=0.55]{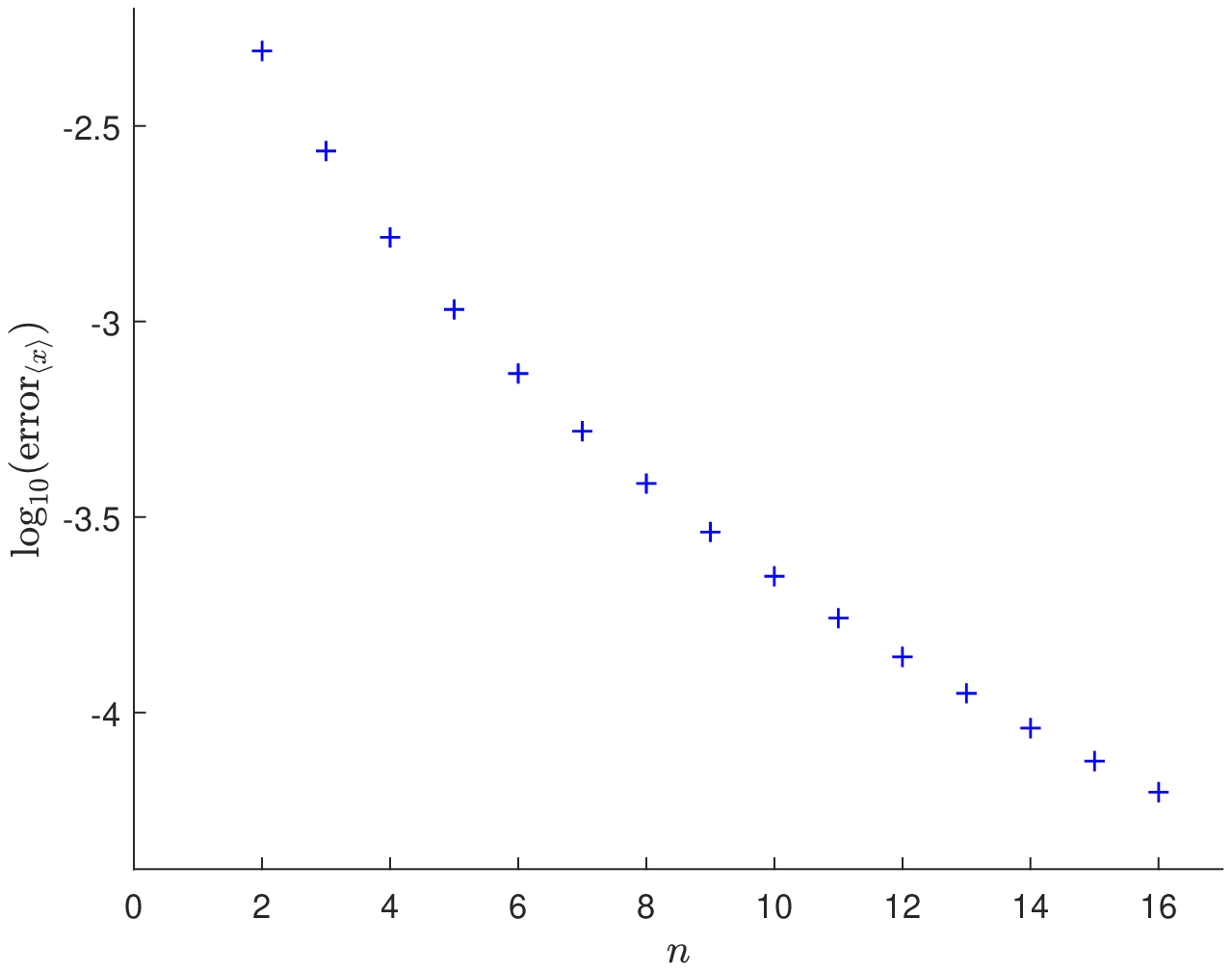}}	
				\caption{The change of the average and the error with the approximation level for the map Eq.~(\ref{formula:5.2.1}) in the new scheme. (a) The average $\langle x\rangle$ becomes more accurate as $n$ increases and converges to $0.3925$ when $n=17$. (b) The dependence of the error of $\langle x\rangle$ on the degree $n$.}			
				\label{graphic:2.1}
			\end{figure}

			Another intermittent map is introduced to validate our method with a slightly different treatment. The Bernoulli shift map~\cite{knight1988deterministic} is modified to
			\begin{eqnarray}{\label{formula:5.2.1}}
			x \mapsto f(x)=\begin{cases}
			f_0(x)=\frac{x}{1-\sqrt[3]{x/4}}&x\in\mathcal{M}_0=[0,\frac{1}{2}]\\
			f_1(x)=2x-1&x\in\mathcal{M}_1=[\frac{1}{2},1] 
			\end{cases},
			\end{eqnarray}
			which is still denoted by $f(x)$ for convenience. Similarly, the Bernoulli shift map has a marginal unstable fixed point at $x=0$ and takes the asymptotic form
			\begin{eqnarray}{\label{formula:5.2.2}}
				f(x)=\sum_{i=0}^{\infty}x(\frac{x}{4})^{i/3}=x+\frac{x^{4/3}}{\sqrt[3]{4}}+\mathcal{O}(x^{5/3}),\,x\to 0.
			\end{eqnarray}
			\begin{figure}[H]	
				\centering
				\includegraphics[scale=0.7]{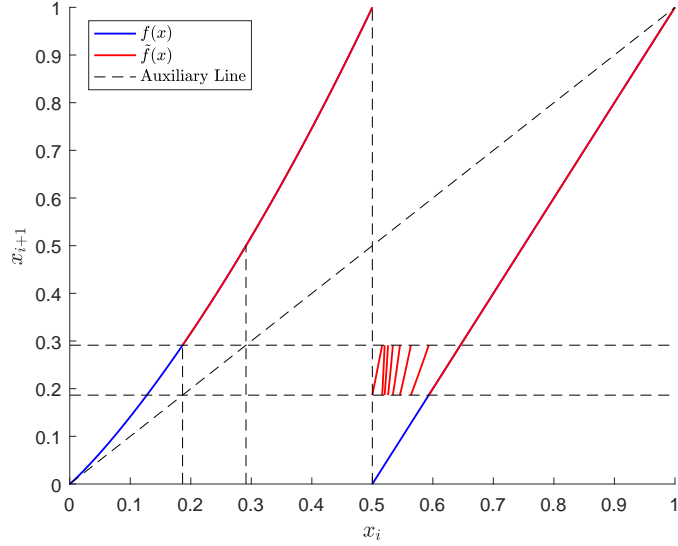}
				\caption{The Bernoulli shift map $f(x)$ (blue line) and the revised map $\tilde{f}(x)$ (red line, $n=8$). The two images coincide on the interval $[f^{-2}_0(q_1),1/2]$ and $(q_2,1]$. Infinite branches of $\tilde{f}(x)$ on $(1/2,q_n]$ are approximated by just one branch $\hat{f}_n(x)$. Also, $\hat{f}_n(\mathcal{M}_n)\subset[f^{-2}_0(q_1),f^{-1}_0(q_1)]$, a range shared by all the $f_m(x)$'s. Here $n$ describes the degree of approximation and the approximation becomes more accurate when $n$ gets larger.}			
				\label{graphic:exam2}
			\end{figure}	 
			As done above, we set $q_1=1/2$ and choose $[0,f^{-2}_0(q_1))$ as the marginal part and $[f^{-2}_0(q_1),1]$ as the hyperbolic. The revised map (Fig.~\ref{graphic:exam2}) and observable $\tilde{a}(x)$ is defined on the hyperbolic part accordingly. The natural measure of $f(x)$ in the interval $(q_1,q_n]$ is approximated as $\rho(x)\approx \omega_0+\omega_1 x$ via cycle expansion and the lowest-order polynomial expansion (see App.~\ref{appendix:7.3}). Being accurate enough, the natural measure of $f(x)$ near $x=0$ is approximately equal to $\rho(x)=\frac{2\omega_0+\omega_1}{2\sqrt[3]{2}}x^{-1/3}+\frac{2\omega_0+\omega_1}{24}+\frac{7\sqrt[3]{2}(2\omega_0+\omega_1)}{432}x^{1/3}+\mathcal{O}(x^{2/3})$. With the current scheme, $\langle x\rangle$ converges to $0.3925$ at $n=17$ (Fig.~\ref{graphic:2.1}(a)), which is closed to the time average $0.3921...$ with the standard deviation $1.238...\times10^{-4}$ and much more precise than $0.32$, the result produced by a direct application of the dynamical zeta function.
	\section{\label{sec:6}summary}
		Maps with marginal orbits produce natural measures with singularities and incur bad convergence in cycle expansions. How to deal with singularities and depict the natural measure correctly is the key to obtain dynamical averages accurately. Essentially, the central idea of this paper is to take advantage of the partial integrability of intermittent systems and analytically estimate the natural measure near the singularities. We divide the phase space into the marginal and the hyperbolic parts in which analytic approximation is used in the marginal part and cycle expansions in the other. As discussed above, we obtained averages of observables very precisely compared to a direct application of the dynamical zeta function. A comb structure is introduced to maintain the natural measure in the hyperbolic part and a redefinition of observables reduces greatly the size of the interval for analytical estimation. Cycle expansions in the hyperbolic part are certainly accelerated due to the removal of singularities.
		
		We test our method on several maps. With a proper analytic computation intermittent near the singularity, we get the averages accurate up to the order of $10^{-4}$, which is far better than a direct application of cycle expansion. With an increase of the approximation level the error decreases steadily. A possible way to further improve the accuracy is to replace the last branch $\hat{f}_n(x)$ in $\mathcal{M}_n$ by a more appropriate nonlinear approximation instead of a simple linear function as in Sect.~\ref{sec:4}. For some other intermittent systems with divergent or non-integrable singularities, however, the computation of $\lambda$ and dynamical averages becomes impossible due to the divergence of natural measure near the singularity.
		
		In this paper, we demonstrate the effectiveness of a new scheme for computing dynamical averages on $1$-dimensional maps with one single singularity. Nevertheless, how to apply the method to $1$-dimensional cases with more singularities or high-dimensional systems or even to flows requires further investigation. In high-dimensional chaotic systems, the natural measure generated on strange attractors with complex structures and fractal dimensions is hard to be estimated analytically. In this case, extra techniques have to be invented to deal with the non-hyperbolic part of the dynamics. Besides, efficient numerical determination of all the shortest periodic orbits in a given nonlinear system is also a major challenge in the application of cycle expansions. Fortunately, good candidates have to be designed to locate UPOs in high-dimensional space~\cite{lan2004variational}.
	\begin{acknowledgements}
		This work was supported by the National Natural Science Foundation of China under Grants No.11775035, and by the Fundamental Research Funds for the Central Universities with contract number 2019XD-A10,  and also by the Key Program of National Natural Science Foundation of China (No. 92067202).
	\end{acknowledgements}
		\bibliography{docbib}

\begin{thebibliography}{30}%
\makeatletter
\providecommand \@ifxundefined [1]{%
 \@ifx{#1\undefined}
}%
\providecommand \@ifnum [1]{%
 \ifnum #1\expandafter \@firstoftwo
 \else \expandafter \@secondoftwo
 \fi
}%
\providecommand \@ifx [1]{%
 \ifx #1\expandafter \@firstoftwo
 \else \expandafter \@secondoftwo
 \fi
}%
\providecommand \natexlab [1]{#1}%
\providecommand \enquote  [1]{``#1''}%
\providecommand \bibnamefont  [1]{#1}%
\providecommand \bibfnamefont [1]{#1}%
\providecommand \citenamefont [1]{#1}%
\providecommand \href@noop [0]{\@secondoftwo}%
\providecommand \href [0]{\begingroup \@sanitize@url \@href}%
\providecommand \@href[1]{\@@startlink{#1}\@@href}%
\providecommand \@@href[1]{\endgroup#1\@@endlink}%
\providecommand \@sanitize@url [0]{\catcode `\\12\catcode `\$12\catcode
  `\&12\catcode `\#12\catcode `\^12\catcode `\_12\catcode `\%12\relax}%
\providecommand \@@startlink[1]{}%
\providecommand \@@endlink[0]{}%
\providecommand \url  [0]{\begingroup\@sanitize@url \@url }%
\providecommand \@url [1]{\endgroup\@href {#1}{\urlprefix }}%
\providecommand \urlprefix  [0]{URL }%
\providecommand \Eprint [0]{\href }%
\providecommand \doibase [0]{https://doi.org/}%
\providecommand \selectlanguage [0]{\@gobble}%
\providecommand \bibinfo  [0]{\@secondoftwo}%
\providecommand \bibfield  [0]{\@secondoftwo}%
\providecommand \translation [1]{[#1]}%
\providecommand \BibitemOpen [0]{}%
\providecommand \bibitemStop [0]{}%
\providecommand \bibitemNoStop [0]{.\EOS\space}%
\providecommand \EOS [0]{\spacefactor3000\relax}%
\providecommand \BibitemShut  [1]{\csname bibitem#1\endcsname}%
\let\auto@bib@innerbib\@empty
\bibitem [{\citenamefont {Lan}(2010)}]{lan2010cycle}%
  \BibitemOpen
  \bibfield  {author} {\bibinfo {author} {\bibfnamefont {Y.}~\bibnamefont
  {Lan}},\ }\bibfield  {title} {\bibinfo {title} {Cycle expansions: From maps
  to turbulence},\ }\href@noop {} {\bibfield  {journal} {\bibinfo  {journal}
  {Commun. Nonlinear Sci.}\ }\textbf {\bibinfo {volume} {15}},\ \bibinfo
  {pages} {502} (\bibinfo {year} {2010})}\BibitemShut {NoStop}%
\bibitem [{\citenamefont {Hao}(1990)}]{hao1990chaos}%
  \BibitemOpen
  \bibfield  {author} {\bibinfo {author} {\bibfnamefont {B.-L.}\ \bibnamefont
  {Hao}},\ }\href@noop {} {\emph {\bibinfo {title} {Chaos II}}}\ (\bibinfo
  {publisher} {World Scientific},\ \bibinfo {year} {1990})\BibitemShut
  {NoStop}%
\bibitem [{\citenamefont {Artuso}\ \emph
  {et~al.}(1990{\natexlab{a}})\citenamefont {Artuso}, \citenamefont {Aurell},\
  and\ \citenamefont {Cvitanovic}}]{artuso1990recycling1}%
  \BibitemOpen
  \bibfield  {author} {\bibinfo {author} {\bibfnamefont {R.}~\bibnamefont
  {Artuso}}, \bibinfo {author} {\bibfnamefont {E.}~\bibnamefont {Aurell}},\
  and\ \bibinfo {author} {\bibfnamefont {P.}~\bibnamefont {Cvitanovic}},\
  }\bibfield  {title} {\bibinfo {title} {Recycling of strange sets: I. cycle
  expansions},\ }\href@noop {} {\bibfield  {journal} {\bibinfo  {journal}
  {Nonlinearity}\ }\textbf {\bibinfo {volume} {3}},\ \bibinfo {pages} {325}
  (\bibinfo {year} {1990}{\natexlab{a}})}\BibitemShut {NoStop}%
\bibitem [{\citenamefont {Cvitanovic}\ \emph {et~al.}(2005)\citenamefont
  {Cvitanovic}, \citenamefont {Artuso}, \citenamefont {Mainieri}, \citenamefont
  {Tanner}, \citenamefont {Vattay}, \citenamefont {Whelan},\ and\ \citenamefont
  {Wirzba}}]{cvitanovic2005chaos}%
  \BibitemOpen
  \bibfield  {author} {\bibinfo {author} {\bibfnamefont {P.}~\bibnamefont
  {Cvitanovic}}, \bibinfo {author} {\bibfnamefont {R.}~\bibnamefont {Artuso}},
  \bibinfo {author} {\bibfnamefont {R.}~\bibnamefont {Mainieri}}, \bibinfo
  {author} {\bibfnamefont {G.}~\bibnamefont {Tanner}}, \bibinfo {author}
  {\bibfnamefont {G.}~\bibnamefont {Vattay}}, \bibinfo {author} {\bibfnamefont
  {N.}~\bibnamefont {Whelan}},\ and\ \bibinfo {author} {\bibfnamefont
  {A.}~\bibnamefont {Wirzba}},\ }\bibfield  {title} {\bibinfo {title} {Chaos:
  classical and quantum},\ }\href@noop {} {\bibfield  {journal} {\bibinfo
  {journal} {ChaosBook. org (Niels Bohr Institute, Copenhagen 2005)}\ }\textbf
  {\bibinfo {volume} {69}},\ \bibinfo {pages} {25} (\bibinfo {year}
  {2005})}\BibitemShut {NoStop}%
\bibitem [{\citenamefont {Auerbach}\ \emph {et~al.}(1987)\citenamefont
  {Auerbach}, \citenamefont {Cvitanovi{\'c}}, \citenamefont {Eckmann},
  \citenamefont {Gunaratne},\ and\ \citenamefont
  {Procaccia}}]{auerbach1987exploring}%
  \BibitemOpen
  \bibfield  {author} {\bibinfo {author} {\bibfnamefont {D.}~\bibnamefont
  {Auerbach}}, \bibinfo {author} {\bibfnamefont {P.}~\bibnamefont
  {Cvitanovi{\'c}}}, \bibinfo {author} {\bibfnamefont {J.-P.}\ \bibnamefont
  {Eckmann}}, \bibinfo {author} {\bibfnamefont {G.}~\bibnamefont {Gunaratne}},\
  and\ \bibinfo {author} {\bibfnamefont {I.}~\bibnamefont {Procaccia}},\
  }\bibfield  {title} {\bibinfo {title} {Exploring chaotic motion through
  periodic orbits},\ }\href@noop {} {\bibfield  {journal} {\bibinfo  {journal}
  {Phys. Rev. Lett.}\ }\textbf {\bibinfo {volume} {58}},\ \bibinfo {pages}
  {2387} (\bibinfo {year} {1987})}\BibitemShut {NoStop}%
\bibitem [{\citenamefont {Cvitanovi{\'c}}\ \emph {et~al.}(1999)\citenamefont
  {Cvitanovi{\'c}}, \citenamefont {S{\o}ndergaard}, \citenamefont {Palla},
  \citenamefont {Vattay},\ and\ \citenamefont
  {Dettmann}}]{cvitanovic1999spectrum}%
  \BibitemOpen
  \bibfield  {author} {\bibinfo {author} {\bibfnamefont {P.}~\bibnamefont
  {Cvitanovi{\'c}}}, \bibinfo {author} {\bibfnamefont {N.}~\bibnamefont
  {S{\o}ndergaard}}, \bibinfo {author} {\bibfnamefont {G.}~\bibnamefont
  {Palla}}, \bibinfo {author} {\bibfnamefont {G.}~\bibnamefont {Vattay}},\ and\
  \bibinfo {author} {\bibfnamefont {C.}~\bibnamefont {Dettmann}},\ }\bibfield
  {title} {\bibinfo {title} {Spectrum of stochastic evolution operators: Local
  matrix representation approach},\ }\href@noop {} {\bibfield  {journal}
  {\bibinfo  {journal} {Phys. Rev. E}\ }\textbf {\bibinfo {volume} {60}},\
  \bibinfo {pages} {3936} (\bibinfo {year} {1999})}\BibitemShut {NoStop}%
\bibitem [{\citenamefont {Cvitanovic}\ and\ \citenamefont
  {Eckhardt}(1991)}]{cvitanovic1991periodic}%
  \BibitemOpen
  \bibfield  {author} {\bibinfo {author} {\bibfnamefont {P.}~\bibnamefont
  {Cvitanovic}}\ and\ \bibinfo {author} {\bibfnamefont {B.}~\bibnamefont
  {Eckhardt}},\ }\bibfield  {title} {\bibinfo {title} {Periodic orbit
  expansions for classical smooth flows},\ }\href@noop {} {\bibfield  {journal}
  {\bibinfo  {journal} {J. Phys. A-Math. Gen.}\ }\textbf {\bibinfo {volume}
  {24}},\ \bibinfo {pages} {L237} (\bibinfo {year} {1991})}\BibitemShut
  {NoStop}%
\bibitem [{\citenamefont {Cvitanovi{\'c}}(1988)}]{cvitanovic1988invariant}%
  \BibitemOpen
  \bibfield  {author} {\bibinfo {author} {\bibfnamefont {P.}~\bibnamefont
  {Cvitanovi{\'c}}},\ }\bibfield  {title} {\bibinfo {title} {Invariant
  measurement of strange sets in terms of cycles},\ }\href@noop {} {\bibfield
  {journal} {\bibinfo  {journal} {Phys. Rev. Lett.}\ }\textbf {\bibinfo
  {volume} {61}},\ \bibinfo {pages} {2729} (\bibinfo {year}
  {1988})}\BibitemShut {NoStop}%
\bibitem [{\citenamefont {Artuso}\ \emph
  {et~al.}(1990{\natexlab{b}})\citenamefont {Artuso}, \citenamefont {Aurell},\
  and\ \citenamefont {Cvitanovic}}]{artuso1990recycling2}%
  \BibitemOpen
  \bibfield  {author} {\bibinfo {author} {\bibfnamefont {R.}~\bibnamefont
  {Artuso}}, \bibinfo {author} {\bibfnamefont {E.}~\bibnamefont {Aurell}},\
  and\ \bibinfo {author} {\bibfnamefont {P.}~\bibnamefont {Cvitanovic}},\
  }\bibfield  {title} {\bibinfo {title} {{Recycling of strange sets: II.
  Applications}},\ }\href@noop {} {\bibfield  {journal} {\bibinfo  {journal}
  {Nonlinearity}\ }\textbf {\bibinfo {volume} {3}},\ \bibinfo {pages} {361}
  (\bibinfo {year} {1990}{\natexlab{b}})}\BibitemShut {NoStop}%
\bibitem [{\citenamefont {Chat{\'e}}\ and\ \citenamefont
  {Manneville}(1987)}]{chate1987transition}%
  \BibitemOpen
  \bibfield  {author} {\bibinfo {author} {\bibfnamefont {H.}~\bibnamefont
  {Chat{\'e}}}\ and\ \bibinfo {author} {\bibfnamefont {P.}~\bibnamefont
  {Manneville}},\ }\bibfield  {title} {\bibinfo {title} {Transition to
  turbulence via spatio-temporal intermittency},\ }\href@noop {} {\bibfield
  {journal} {\bibinfo  {journal} {Phys. Rev. Lett.}\ }\textbf {\bibinfo
  {volume} {58}},\ \bibinfo {pages} {112} (\bibinfo {year} {1987})}\BibitemShut
  {NoStop}%
\bibitem [{\citenamefont {Chate}(1994)}]{chate1994spatiotemporal}%
  \BibitemOpen
  \bibfield  {author} {\bibinfo {author} {\bibfnamefont {H.}~\bibnamefont
  {Chate}},\ }\bibfield  {title} {\bibinfo {title} {Spatiotemporal
  intermittency regimes of the one-dimensional complex ginzburg-landau
  equation},\ }\href@noop {} {\bibfield  {journal} {\bibinfo  {journal}
  {Nonlinearity}\ }\textbf {\bibinfo {volume} {7}},\ \bibinfo {pages} {185}
  (\bibinfo {year} {1994})}\BibitemShut {NoStop}%
\bibitem [{\citenamefont {Manneville}\ and\ \citenamefont
  {Pomeau}(1979)}]{manneville1979intermittency}%
  \BibitemOpen
  \bibfield  {author} {\bibinfo {author} {\bibfnamefont {P.}~\bibnamefont
  {Manneville}}\ and\ \bibinfo {author} {\bibfnamefont {Y.}~\bibnamefont
  {Pomeau}},\ }\bibfield  {title} {\bibinfo {title} {Intermittency and the
  lorenz model},\ }\href@noop {} {\bibfield  {journal} {\bibinfo  {journal}
  {Phys. Lett. A}\ }\textbf {\bibinfo {volume} {75}},\ \bibinfo {pages} {1}
  (\bibinfo {year} {1979})}\BibitemShut {NoStop}%
\bibitem [{\citenamefont {Pomeau}\ and\ \citenamefont
  {Manneville}(1980)}]{pomeau1980intermittent}%
  \BibitemOpen
  \bibfield  {author} {\bibinfo {author} {\bibfnamefont {Y.}~\bibnamefont
  {Pomeau}}\ and\ \bibinfo {author} {\bibfnamefont {P.}~\bibnamefont
  {Manneville}},\ }\bibfield  {title} {\bibinfo {title} {Intermittent
  transition to turbulence in dissipative dynamical systems},\ }\href@noop {}
  {\bibfield  {journal} {\bibinfo  {journal} {Commun. Math. Phys.}\ }\textbf
  {\bibinfo {volume} {74}},\ \bibinfo {pages} {189} (\bibinfo {year}
  {1980})}\BibitemShut {NoStop}%
\bibitem [{\citenamefont {Devaney}(1989)}]{devaney1989introduction}%
  \BibitemOpen
  \bibfield  {author} {\bibinfo {author} {\bibfnamefont {R.~L.}\ \bibnamefont
  {Devaney}},\ }\href@noop {} {\emph {\bibinfo {title} {An introduction to
  chaotic dynamical systems}}}\ (\bibinfo  {publisher} {Chapman and Hall/CRC},\
  \bibinfo {year} {1989})\BibitemShut {NoStop}%
\bibitem [{\citenamefont {Tanner}\ and\ \citenamefont
  {Wintgen}(1995)}]{tanner1995semiclassical}%
  \BibitemOpen
  \bibfield  {author} {\bibinfo {author} {\bibfnamefont {G.}~\bibnamefont
  {Tanner}}\ and\ \bibinfo {author} {\bibfnamefont {D.}~\bibnamefont
  {Wintgen}},\ }\bibfield  {title} {\bibinfo {title} {Semiclassical
  quantization of intermittency in helium},\ }\href@noop {} {\bibfield
  {journal} {\bibinfo  {journal} {Phys. Rev. Lett.}\ }\textbf {\bibinfo
  {volume} {75}},\ \bibinfo {pages} {2928} (\bibinfo {year}
  {1995})}\BibitemShut {NoStop}%
\bibitem [{\citenamefont {Tanner}\ \emph {et~al.}(1996)\citenamefont {Tanner},
  \citenamefont {Hansen},\ and\ \citenamefont
  {Main}}]{tanner1996semiclassical}%
  \BibitemOpen
  \bibfield  {author} {\bibinfo {author} {\bibfnamefont {G.}~\bibnamefont
  {Tanner}}, \bibinfo {author} {\bibfnamefont {K.~T.}\ \bibnamefont {Hansen}},\
  and\ \bibinfo {author} {\bibfnamefont {J.}~\bibnamefont {Main}},\ }\bibfield
  {title} {\bibinfo {title} {The semiclassical resonance spectrum of hydrogen
  in a constant magnetic field},\ }\href@noop {} {\bibfield  {journal}
  {\bibinfo  {journal} {Nonlinearity}\ }\textbf {\bibinfo {volume} {9}},\
  \bibinfo {pages} {1641} (\bibinfo {year} {1996})}\BibitemShut {NoStop}%
\bibitem [{\citenamefont {Tanner}(1997)}]{tanner1997chaotic}%
  \BibitemOpen
  \bibfield  {author} {\bibinfo {author} {\bibfnamefont {G.}~\bibnamefont
  {Tanner}},\ }\bibfield  {title} {\bibinfo {title} {How chaotic is the stadium
  billiard? a semiclassical analysis},\ }\href@noop {} {\bibfield  {journal}
  {\bibinfo  {journal} {J. Phys. A-Math. Gen.}\ }\textbf {\bibinfo {volume}
  {30}},\ \bibinfo {pages} {2863} (\bibinfo {year} {1997})}\BibitemShut
  {NoStop}%
\bibitem [{\citenamefont {Dettmann}\ and\ \citenamefont
  {Morriss}(1997)}]{dettmann1997stability}%
  \BibitemOpen
  \bibfield  {author} {\bibinfo {author} {\bibfnamefont {C.}~\bibnamefont
  {Dettmann}}\ and\ \bibinfo {author} {\bibfnamefont {G.}~\bibnamefont
  {Morriss}},\ }\bibfield  {title} {\bibinfo {title} {Stability ordering of
  cycle expansions},\ }\href@noop {} {\bibfield  {journal} {\bibinfo  {journal}
  {Phys. Rev. Lett.}\ }\textbf {\bibinfo {volume} {78}},\ \bibinfo {pages}
  {4201} (\bibinfo {year} {1997})}\BibitemShut {NoStop}%
\bibitem [{\citenamefont {Artuso}\ \emph {et~al.}(2003)\citenamefont {Artuso},
  \citenamefont {Cvitanovi{\'c}},\ and\ \citenamefont
  {Tanner}}]{artuso2003cycle}%
  \BibitemOpen
  \bibfield  {author} {\bibinfo {author} {\bibfnamefont {R.}~\bibnamefont
  {Artuso}}, \bibinfo {author} {\bibfnamefont {P.}~\bibnamefont
  {Cvitanovi{\'c}}},\ and\ \bibinfo {author} {\bibfnamefont {G.}~\bibnamefont
  {Tanner}},\ }\bibfield  {title} {\bibinfo {title} {Cycle expansions for
  intermittent maps},\ }\href@noop {} {\bibfield  {journal} {\bibinfo
  {journal} {Prog. Theor. Phys. Supp.}\ }\textbf {\bibinfo {volume} {150}},\
  \bibinfo {pages} {1} (\bibinfo {year} {2003})}\BibitemShut {NoStop}%
\bibitem [{\citenamefont {Prellberg}\ and\ \citenamefont
  {Slawny}(1992)}]{prellberg1992maps}%
  \BibitemOpen
  \bibfield  {author} {\bibinfo {author} {\bibfnamefont {T.}~\bibnamefont
  {Prellberg}}\ and\ \bibinfo {author} {\bibfnamefont {J.}~\bibnamefont
  {Slawny}},\ }\bibfield  {title} {\bibinfo {title} {Maps of intervals with
  indifferent fixed points: thermodynamic formalism and phase transitions},\
  }\href@noop {} {\bibfield  {journal} {\bibinfo  {journal} {J. Stat. Phys.}\
  }\textbf {\bibinfo {volume} {66}},\ \bibinfo {pages} {503} (\bibinfo {year}
  {1992})}\BibitemShut {NoStop}%
\bibitem [{\citenamefont {Cvitanovi{\'c}}(1991)}]{cvitanovic1991periodic2}%
  \BibitemOpen
  \bibfield  {author} {\bibinfo {author} {\bibfnamefont {P.}~\bibnamefont
  {Cvitanovi{\'c}}},\ }\bibfield  {title} {\bibinfo {title} {Periodic orbits as
  the skeleton of classical and quantum chaos},\ }\href@noop {} {\bibfield
  {journal} {\bibinfo  {journal} {Physica D}\ }\textbf {\bibinfo {volume}
  {51}},\ \bibinfo {pages} {138} (\bibinfo {year} {1991})}\BibitemShut
  {NoStop}%
\bibitem [{\citenamefont {Ruelle}(2004)}]{ruelle2004thermodynamic}%
  \BibitemOpen
  \bibfield  {author} {\bibinfo {author} {\bibfnamefont {D.}~\bibnamefont
  {Ruelle}},\ }\href@noop {} {\emph {\bibinfo {title} {Thermodynamic formalism:
  the mathematical structure of equilibrium statistical mechanics}}}\ (\bibinfo
   {publisher} {Cambridge University Press},\ \bibinfo {year}
  {2004})\BibitemShut {NoStop}%
\bibitem [{\citenamefont {Rugh}(1992)}]{rugh1992correlation}%
  \BibitemOpen
  \bibfield  {author} {\bibinfo {author} {\bibfnamefont {H.~H.}\ \bibnamefont
  {Rugh}},\ }\bibfield  {title} {\bibinfo {title} {The correlation spectrum for
  hyperbolic analytic maps},\ }\href@noop {} {\bibfield  {journal} {\bibinfo
  {journal} {Nonlinearity}\ }\textbf {\bibinfo {volume} {5}},\ \bibinfo {pages}
  {1237} (\bibinfo {year} {1992})}\BibitemShut {NoStop}%
\bibitem [{\citenamefont {Sinai}(1968)}]{sinai1968construction}%
  \BibitemOpen
  \bibfield  {author} {\bibinfo {author} {\bibfnamefont {Y.~G.}\ \bibnamefont
  {Sinai}},\ }\bibfield  {title} {\bibinfo {title} {Construction of markov
  partitions},\ }\href@noop {} {\bibfield  {journal} {\bibinfo  {journal}
  {Funct. Anal. Appl.+}\ }\textbf {\bibinfo {volume} {2}},\ \bibinfo {pages}
  {245} (\bibinfo {year} {1968})}\BibitemShut {NoStop}%
\bibitem [{\citenamefont {Kitchens}(2012)}]{kitchens2012symbolic}%
  \BibitemOpen
  \bibfield  {author} {\bibinfo {author} {\bibfnamefont {B.~P.}\ \bibnamefont
  {Kitchens}},\ }\href@noop {} {\emph {\bibinfo {title} {Symbolic dynamics:
  one-sided, two-sided and countable state Markov shifts}}}\ (\bibinfo
  {publisher} {Springer Science \& Business Media},\ \bibinfo {year}
  {2012})\BibitemShut {NoStop}%
\bibitem [{\citenamefont {Collet}\ and\ \citenamefont
  {Eckmann}(2009)}]{collet2009iterated}%
  \BibitemOpen
  \bibfield  {author} {\bibinfo {author} {\bibfnamefont {P.}~\bibnamefont
  {Collet}}\ and\ \bibinfo {author} {\bibfnamefont {J.-P.}\ \bibnamefont
  {Eckmann}},\ }\href@noop {} {\emph {\bibinfo {title} {Iterated maps on the
  interval as dynamical systems}}}\ (\bibinfo  {publisher} {Springer Science \&
  Business Media},\ \bibinfo {year} {2009})\BibitemShut {NoStop}%
\bibitem [{\citenamefont {Metropolis}\ \emph {et~al.}(1973)\citenamefont
  {Metropolis}, \citenamefont {Stein},\ and\ \citenamefont
  {Stein}}]{metropolis1973finite}%
  \BibitemOpen
  \bibfield  {author} {\bibinfo {author} {\bibfnamefont {N.}~\bibnamefont
  {Metropolis}}, \bibinfo {author} {\bibfnamefont {M.}~\bibnamefont {Stein}},\
  and\ \bibinfo {author} {\bibfnamefont {P.}~\bibnamefont {Stein}},\ }\bibfield
   {title} {\bibinfo {title} {On finite limit sets for transformations on the
  unit interval},\ }\href@noop {} {\bibfield  {journal} {\bibinfo  {journal}
  {J. Comb. Theory A}\ }\textbf {\bibinfo {volume} {15}},\ \bibinfo {pages}
  {25} (\bibinfo {year} {1973})}\BibitemShut {NoStop}%
\bibitem [{\citenamefont {Knight}(1988)}]{knight1988deterministic}%
  \BibitemOpen
  \bibfield  {author} {\bibinfo {author} {\bibfnamefont {P.}~\bibnamefont
  {Knight}},\ }\href@noop {} {\bibinfo {title} {Deterministic chaos: An
  introduction}} (\bibinfo {year} {1988})\BibitemShut {NoStop}%
\bibitem [{\citenamefont {Lan}\ and\ \citenamefont
  {Cvitanovi{\'c}}(2004)}]{lan2004variational}%
  \BibitemOpen
  \bibfield  {author} {\bibinfo {author} {\bibfnamefont {Y.}~\bibnamefont
  {Lan}}\ and\ \bibinfo {author} {\bibfnamefont {P.}~\bibnamefont
  {Cvitanovi{\'c}}},\ }\bibfield  {title} {\bibinfo {title} {Variational method
  for finding periodic orbits in a general flow},\ }\href@noop {} {\bibfield
  {journal} {\bibinfo  {journal} {Phys. Rev. E}\ }\textbf {\bibinfo {volume}
  {69}},\ \bibinfo {pages} {016217} (\bibinfo {year} {2004})}\BibitemShut
  {NoStop}%
\bibitem [{\citenamefont {Legendre}(1805)}]{legendre1805new}%
  \BibitemOpen
  \bibfield  {author} {\bibinfo {author} {\bibfnamefont {A.}~\bibnamefont
  {Legendre}},\ }\bibfield  {title} {\bibinfo {title} {New methods for the
  determination of orbits of comets},\ }\href@noop {} {\bibfield  {journal}
  {\bibinfo  {journal} {Courcier, Paris}\ } (\bibinfo {year}
  {1805})}\BibitemShut {NoStop}%
\end{thebibliography}%
	\appendixpage
	\appendix{\label{appendix:1}}
		\subsection{\label{appendix:7.1}Justification of Eq.~(\ref{formula:3.1.2})}
			Denote the natural measure of the original map with $\rho(x)$, and the revised map with $\tilde{\rho}(x)$. Obviously, the natural measure satisfies the relation \(\rho(x)=\mathcal{L}^n\rho(x)\), which means it is invariant under system evolution. For the map $f(x)$, we have
			\begin{small}
				\begin{eqnarray}{\label{formula:app.1.1}}
				\rho(x_1)=\frac{\rho(x_{11})}{|f_1'(x_{11})|}+\frac{\rho(x_{01})}{|f_0'(x_{01})|},x_1\in(1/2,1]
				\end{eqnarray}
			\end{small}
			where $x_{11}$ and $x_{01}$ denote the preimages of $x_1$ on branch $f_1$ and $f_0$ respectively. Next, by replacing \(\rho(x_{01})\) with the natural measure on its preimages, $\rho(x_{01})=\frac{\rho(x_{101})}{|f_1'(x_{101})|}+\frac{\rho(x_{001})}{|f_0'(x_{001})|}$, we have
			\begin{small}
				\begin{eqnarray}{\label{formula:app.1.2}}
				\rho(x_1)=\frac{\rho(x_{11})}{|f_1'(x_{11})|}+\frac{\rho(x_{101})}{|(f_0f_1)'(x_{101})|}+\frac{\rho(x_{001})}{|(f^2_0)'(x_{001})|}.
				\end{eqnarray}
			\end{small}
			Keep doing this, we finally arrive at
			\begin{small}
				\begin{eqnarray}{\label{formula:app.1.3}}
				\rho(x_1)=\frac{\rho(x_{11})}{|f_1'(x_{11})|}+\sum_{m=1}^{\infty}\frac{\rho(x_{10^m1})}{|(f^m_0f_1)'(x_{10^m1})|}.
				\end{eqnarray}
			\end{small}
			The revised map $\tilde{f}(x)$ gives a similar equation for the measure $\tilde{\rho}(x)$
			\begin{small}
				\begin{eqnarray}{\label{formula:app.1.4}}
				\tilde{\rho}(x_1)=\frac{\tilde{\rho}(x_{11})}{|f_1'(x_{11})|}+\sum_{m=1}^{\infty}\frac{\tilde{\rho}(x_{10^m1})}{|(f^m_0f_1)'(x_{10^m1})|},
				\end{eqnarray}
			\end{small}
			which is obviously satisfied by the measure \(\rho(x)\). The uniqueness of the invariant measure indicates that \(\rho(x)=\lambda\tilde{\rho}(x)\) in the hyperbolic part, where $\lambda$ is a constant with $\lambda\in(0,1)$. The equations above are enough to determine the functional form of $\rho(x)$ and $\tilde{\rho}(x)$. It is worth mentioning that we still have \(\rho(x)\approx\lambda\tilde{\rho}(x)\) in the approximation with an arbitrarily selected $n$ defined in Sect.~\ref{sec:4}. The larger the $n$, the more accurate the approximation.
			
		\subsection{\label{appendix:7.2}Justification of Eq.~(\ref{formula:3.2.2})}
			We expand the natural measure on $[0,1/2]$ of the map $f(x)$ similar to Eq.~(\ref{formula:app.1.3}) above. The result is
			\begin{small}
				\begin{eqnarray}{\label{formula:app.2.1}}
				\rho(x_0)=\sum_{m=2}^{\infty}\frac{\rho(x_{10^{m-1}})}{|(f^{m-2}_0f_1)'(x_{10^{m-1}})|},x_0\in[0,1/2].
				\end{eqnarray}
			\end{small}
			It’s easy to see that the right-hand side of the equation involves only points and natural measure defined on the interval $[1/2,1]$, which indicates that its direct evaluation on $[0,1/2]$ could be avoided.
			
			The average on $[0,1/2]$ satisfies
			\begin{small}
				\begin{eqnarray}{\label{formula:app.2.2}}
					\begin{split}
						&\hspace{5mm}\int_{0}^{1/2}\rho(x_0)a(x_0)dx_0\\
						&=\sum_{m=2}^{\infty}\int_{0}^{1/2}\frac{\rho(x_{10^{m-1}})}{|(f^{m-2}_0f_1)'(x_{10^{m-1}})|}a(x_0)dx_0\\
						&=\sum_{m=2}^{\infty}\int_{q_m}^{1}\rho(x_{10^{m-1}})a((f^{m-2}_0f_1)(x_{10^{m-1}}))dx_{10^{m-1}}\\
						&=\sum_{m=2}^{\infty}\int_{\mathcal{M}_m}\rho(x_{10^{m-1}})\sum_{n=2}^{m}a((f^{n-2}_0f_1)(x_{10^{m-1}}))dx_{10^{m-1}}\\
						&=\sum_{m=2}^{\infty}\int_{\mathcal{M}_m}\rho(x)\sum_{n=2}^{m}a((f^{n-2}_0f_1)(x))dx
						\,.
					\end{split}
				\end{eqnarray}
			\end{small}
			Recalling the definition of $\tilde{a}(x)$, we get
			\begin{small}
				\begin{eqnarray}{\label{formula:app.2.3}}
				\begin{split}
				<a>&=\int_{0}^{1}\rho(x)a(x)dx\\
				&=\int_{0}^{1/2}\rho(x)a(x)dx+\int_{1/2}^{1}\rho(x)a(x)dx\\
				&=\int_{1/2}^{1}\rho(x)\tilde{a}(x)dx\\
				&=\int_{1/2}^{1}\lambda\tilde{\rho}(x)\tilde{a}(x)dx\\
				&=\lambda<\tilde{a}>.
				\end{split}.
				\end{eqnarray}
			\end{small}
		\subsection{\label{appendix:7.3}Method to Derive Approximate Natural Measure (Sect.~\ref{subsec:5.1}, Sect.~\ref{subsec:5.2})}
			For a 1-dimensional map $f(x)$ which takes the form of Eq.~(\ref{Formula:2.2}) near the marginally unstable fixed point $x=0$, its natural measure takes the form $\rho\sim\frac{1}{x^s}$ near the point. Therefore we can reasonably assume that the expression of the natural measure is
			\begin{small}\begin{eqnarray}{\label{formula:app.3.1.1}}
				\rho(x)=\sum_{i=0}^{\infty}\sum_{j>-s}^{\infty}\alpha_{ij}x^{(i-1)s+j},\,x\to0,
			\end{eqnarray}\end{small}
			where $\alpha_{ij}$ equals zero when $j$ is a multiple of $s$ to avoid repetition. We select an arbitrary $n$ defined in Sect.~\ref{sec:4} and all the following ``$\approx$'' are due to this approximation. The natural measure in $\mathcal{M}_n$ can be expanded as Legendre polynomials \cite{legendre1805new}
			\begin{small}\begin{align}{\label{formula:Legendre}}
				\begin{split}
				&\rho(x)=\sum_{l=0}^{\infty}c_lP_l(\phi(x)),x\in\mathcal{M}_n=(q_n,1],\\
				&\mathrm{with}\quad\phi(x)=\frac{2}{1-q_n}x-\frac{1+q_n}{1-q_n},\\
				\end{split}
			\end{align}\end{small}
			where $\phi(x)$ describes the shift and scaling of the interval $\mathcal{M}_n$ being mapped to $(-1,1]$. The coefficient $c_l$ can be evaluated through
			\begin{small}\begin{align}{\label{formula:c_l}}
				\begin{split}
				c_l&=\frac{2l+1}{2}\int_{q_n}^{1}\rho(x)P_l(\phi(x))\phi'(x)dx\\
				&\approx\frac{2l+1}{2}\int_{q_n}^{1}\lambda\tilde{\rho}(x)P_l(\phi(x))\phi'(x)dx=\lambda\tilde{c}_l,\\
				&\mathrm{with}\quad P_0(x)=1,P_1(x)=x,P_2(x)=\frac{3}{2}x^2-\frac{1}{2},...\\
				\end{split}
				\end{align}\end{small}
			In the above equation, $\frac{2l+1}{2}P_l(\phi(x))\phi'(x)$ is regarded as an observable and $c_l$ is computed through cycle expansion as an average defined by Eq.~(\ref{Formula:phaseaverage}). Substituting Eq.~(\ref{formula:app.3.1.1}), (\ref{formula:Legendre}) and (\ref{formula:c_l}) into Eq.~(\ref{formula:4.2}), we have
			\begin{small}\begin{align}{\label{formula:app.3.1.2}}
				\rho(x)=&\frac{\rho(x_0)}{|f_0'(x_0)|}+\frac{\rho(x_1)}{|f_1'(x_1)|}\nonumber\\
				\approx&\frac{\sum_{i=0}^{\infty}\sum_{j>-s}^{\infty}\alpha_{ij}x^{(i-1)s+j}}{|f_0'(x_0)|}+\frac{\sum_{l=0}^{\infty}\lambda\tilde{c}_lP_l(\phi(x_1))}{|f_1'(x_1)|}\nonumber\\
				\approx&\sum_{i=0}^{\infty}\sum_{j>-s}^{\infty}\alpha_{ij}x^{(i-1)s+j},x\to0,
			\end{align}\end{small}
			where $x_0=f^{-1}_0(x)$, $x_1=f^{-1}_1(x)$, $f_0'$ and $f_1'$ can all be expanded as series expansions of $x$. Therefore, the coefficients $\alpha_{ij}$ can be rewritten as $\alpha_{ij}=\sum_{l=0}^{\infty}\lambda A_{ijl}\tilde{c}_l$, a linear combination of $\tilde{c}_l$ determined by comparing the coefficients of each order of $x$. In practical computation, the more accurate $\rho(x)$ we want, the higher order Eq.~(\ref{formula:app.3.1.1}) needs to be expanded, {\em e.g.}, we need to expand Eq.~(\ref{formula:app.3.1.1}) to $\mathcal{O}(x)$ to determine the coefficients of $\rho(x)$ in Sect.~\ref{subsec:5.1} up to $\mathcal{O}(x^{1/2})$.
		\subsection{\label{appendix:lambda}Method to obtain $\lambda$ (Sect.~\ref{subsec:3.1})}
			Actually, the coefficients of $\rho(x)$ we get in App.~\ref{appendix:7.3} contain the undetermined constant $\lambda$ which needs to be fixed. Fortunately under the necessary approximation (an arbitrarily selected $n$ described in Sect.~\ref{sec:4}), cycle expansions can also provide what we need. Also for the map $f(x)$, we introduce $\hat{\sigma}$ and $\hat{\mu}$ to denote the integral of $\rho(x)$ on different intervals respectively ($\hat{\tilde{\sigma}}$ and $\hat{\tilde{\mu}}$ for $\tilde{\rho}(x)$). We define them as 
			\begin{small}
				\begin{eqnarray}{\label{formula:sigma and mu definition}}
					\begin{cases}
						\hat{\sigma}=&\int_{0}^{1}\rho(x)dx\\
						\hat{\sigma}_h=&\int_{q_1}^{1}\rho(x)dx\\
						\hat{\sigma}_0=&\int_{0}^{f_1(q_n)}\rho(x)dx\\
						\hat{\sigma}_i=&\int_{f_1(q_{n-i+1})}^{f_1(q_{n-i})}\rho(x)dx\\
						\hat{\mu}_{i}=&\int_{q_{n-i}}^{1}\rho(x)dx\\
					\end{cases},i=1,2,...,n-2,
				\end{eqnarray}
			\end{small}	
			where $q_i$ and $f_1$ are both defined as in Sect.~\ref{subsec:3.1} and the subscript $0$ and $h$ denote analytic approximation region and hyperbolic region respectively. Obviously we get
			\begin{eqnarray}{\label{formula:equation of density transpotion}}
				\begin{cases}
					\hat{\sigma}=\hat{\sigma}_0+\sum_{i}\hat{\sigma}_i+\hat{\sigma}_h=1\\
					\hat{\tilde{\sigma}}=\hat{\tilde{\sigma}}_h=1,\,\hat{\tilde{\sigma}}_0=\hat{\tilde{\sigma}}_i=0
				\end{cases},i=1,2,...,n-2.
			\end{eqnarray}
			According to the invariance of the density, we have
			\begin{align}{\label{formula:invariance of density transpotion}}
				&\int_{0}^{f_1(q_{n-i})}\rho(x)dx=\int_{0}^{f_1(q_{n-i+1})}\rho(x)dx+\int_{q_{n-i}}^{1}\rho(x)dx\nonumber\\
				\Rightarrow&\,\int_{f_1(q_{n-i+1})}^{f_1(q_{n-i})}\rho(x)dx=\int_{q_{n-i}}^{1}\rho(x)dx\nonumber\\
				\Rightarrow&\,\hat{\sigma}_i=\hat{\mu}_i,\,i=1,2,...,n-2
			\end{align}
			From \(\rho(x)=\lambda\tilde{\rho}(x)\) and $\hat{\tilde{\sigma}}=\hat{\tilde{\sigma}}_h=1$ in Eq.~(\ref{formula:equation of density transpotion}), we get
			\begin{align}{\label{formula:sigma_h}}
				\hat{\sigma}_h=\lambda\hat{\tilde{\sigma}}_h=\lambda.
			\end{align}
			Substitute Eq.~(\ref{formula:sigma_h}) and (\ref{formula:invariance of density transpotion}) into the first equation of Eq.~(\ref{formula:equation of density transpotion}), we have
			\begin{align}{\label{formula:sigma}}
				&\hat{\sigma}=\hat{\sigma}_0+\sum_{i}\hat{\sigma}_i+\hat{\sigma}_h=1\nonumber\\
				\Rightarrow&\int_{0}^{f_1(q_n)}\rho(x)dx_0+\sum_{i}\lambda\hat{\tilde{\mu}}_i+\lambda=1\nonumber\\
				\Rightarrow&\lambda=\frac{1}{\int_{0}^{f_1(q_n)}\sum_{i=0}^{\infty}\sum_{j>-s}^{\infty}\sum_{l=0}^{\infty}A_{ijl}\tilde{c}_lx_0^{(i-1)s+j}dx_0+1+\sum_{i}\hat{\tilde{\mu}}_i},
			\end{align}
			where both $\hat{\tilde{\mu}}_i=\int_{q_{n-i}}^{1}\tilde{\rho}(x)dx$ and $\tilde{c}_l$ can be obtained through cycle expansions. $A_{ijl}$ are obtained by analytic approximation in App.~\ref{appendix:7.3}. Further derivation of the dynamical average $\langle a\rangle$ is carried out as 
			\begin{align}{\label{formula:final observable a}}
				\begin{split}
				\langle a\rangle=&\int_{0}^{f_1(q_n)}+\int_{f_1(q_{n})}^{q_1}+\int_{q_1}^{1}\rho(x)a(x)dx\\
					=&\int_{0}^{f_1(q_n)}\sum_{i=0}^{\infty}\sum_{j>-s}^{\infty}\alpha_{ij}x^{(i-1)s+j}a(x)dx\\
					&+\sum_{i=2}^{n-1}\int_{q_{i}}^{q_{i+1}}\sum_{j=2}^{i}\lambda\tilde{\rho}(x)a((f_0^{j-2}f_1)(x))dx+\int_{q_n}^{1}\sum_k\lambda\tilde{\rho}(x)a((f_0^kf_1)(x))dx\\
					&+\int_{q_1}^{1}\lambda\tilde{\rho}(x)a(x)dx\\
					=&\langle a\rangle_0+\lambda\langle\tilde{a}\rangle,
				\end{split}
			\end{align}
			where $k$ denotes all positive integers such that $(f^k_0f_1)(x)\in(f_1(q_n),q_1]$. Thus with the approximation in Sect.~\ref{sec:4}, we change the form of the observable $\tilde{a}(x)$ in $\mathcal{M}_n$ to $a(x)+\sum_{k}^{}a((f_0^{k}f_1)(x))$ accordingly. Finally, the average $\langle a\rangle$ we need can be obtained from the analytic approximation $\langle a\rangle_0$ in the neighborhood of $x=0$ and the average of the observable $\langle\tilde{a}\rangle$ computed on the hyperbolic part through Eq.~(\ref{formula:final average}).
\end{document}